\documentclass[superscriptaddress,pra,amsmath,amssymb,eufrak,aps,floatfix,twocolumn,notitlepage,longbibliography]{revtex4-1}

\usepackage{lipsum}
\usepackage{braket}
\usepackage{graphicx}
\usepackage{dcolumn}
\usepackage{enumitem}
\usepackage{bm}
\usepackage{amsmath}
\usepackage{amssymb}
\usepackage{mathtools}
\usepackage{subfigure}
\usepackage{color}
\usepackage{xcolor}
\usepackage{verbatim}
\usepackage{float}
\usepackage{appendix}
\usepackage{bbold}
\usepackage{mathrsfs}

\newcommand{\abs}[1]{\lvert #1 \rvert} 
\newcommand{\ev}[1]{\langle #1 \rangle} 
\newcommand{\td}[1]{\tilde{#1}} 
\newcommand{\mbf}[1]{\mathbf{#1}} 
\newcommand{\hS}[0]{\hat{s}} 
\newcommand{\hSS}[0]{\hat{\mathfrak{s}}} 
\newcommand{\cSS}[0]{\mathfrak{s}}
\newcommand{\mcX}[0]{\mathcal{X}} 
\newcommand{\mcY}[0]{\mathcal{Y}} 
\newcommand{\mcZ}[0]{\mathcal{Z}} 
\newcommand{\mfJ}[0]{\mathfrak{J}} 
\makeatletter
\newcommand*{\balancecolsandclearpage}{%
  \close@column@grid
  \cleardoublepage
  \twocolumngrid
}

\usepackage{upgreek}
\usepackage{physics}
\usepackage{isomath}
\usepackage{placeins}
\usepackage{hhline}
\usepackage{hyperref}
\usepackage{cancel}

\begin{document}

\preprint{APS/123-QED}

\title{Simulating  dynamical phases of chiral  $p+ i p$ superconductors with a trapped ion magnet }

\author{Athreya Shankar}
\email{athreya.shankar@uibk.ac.at}
\affiliation{Institute for Theoretical Physics, University of Innsbruck, Innsbruck, Austria}
\affiliation{Institute for Quantum Optics and Quantum Information of the Austrian Academy of Sciences, Innsbruck, Austria}
\author{Emil A.  Yuzbashyan}
\affiliation{Center
for Materials Theory, Rutgers University, Piscataway, New Jersey 08854, USA}
\author{Victor Gurarie}
\affiliation{Department of Physics, University of Colorado, Boulder, CO 80309}
\affiliation{Center for Theory of Quantum Matter, University of Colorado, Boulder, CO 80309}
\author{Peter Zoller}
\affiliation{Institute for Theoretical Physics, University of Innsbruck, Innsbruck, Austria}
\affiliation{Institute for Quantum Optics and Quantum Information of the Austrian Academy of Sciences, Innsbruck, Austria}
\author{John J. Bollinger}
\affiliation{National Institute of Standards and Technology, Boulder, CO 80309}
\author{Ana Maria Rey}
\email{arey@jilau1.colorado.edu}
\affiliation{JILA, National Institute of Standards and Technology,and Department of Physics, University of Colorado, Boulder, CO 80309}
\affiliation{Center for Theory of Quantum Matter, University of Colorado, Boulder, CO 80309}

\date{\today}

\begin{abstract}
Two-dimensional  $p+ i p$  superconductors and superfluids  are systems that  feature  chiral behavior  emerging from  the Cooper pairing  of electrons  or neutral fermionic atoms with non-zero angular momentum. Their realization has been a longstanding goal because they offer great potential utility for quantum computation and memory. However,  they have so far eluded  experimental observation both in  solid state systems as well as in  ultracold quantum gases. Here, we propose to leverage the tremendous control offered by rotating two-dimensional  trapped-ion crystals in a Penning trap  to simulate the dynamical phases of two-dimensional $p+ip$ superfluids. This is accomplished by  mapping the presence or absence of a Cooper pair into an effective spin-1/2 system encoded in the ions' electronic levels. We show how to infer the topological properties of the dynamical phases, and discuss the role of beyond mean-field corrections. More broadly, our work opens the door to use trapped ion systems to explore exotic models of topological superconductivity and also paves the way to generate and manipulate skyrmionic spin textures in these platforms.
\end{abstract}

\maketitle


\begin{figure*}[!tb]
    \centering
    \includegraphics[width=\textwidth]{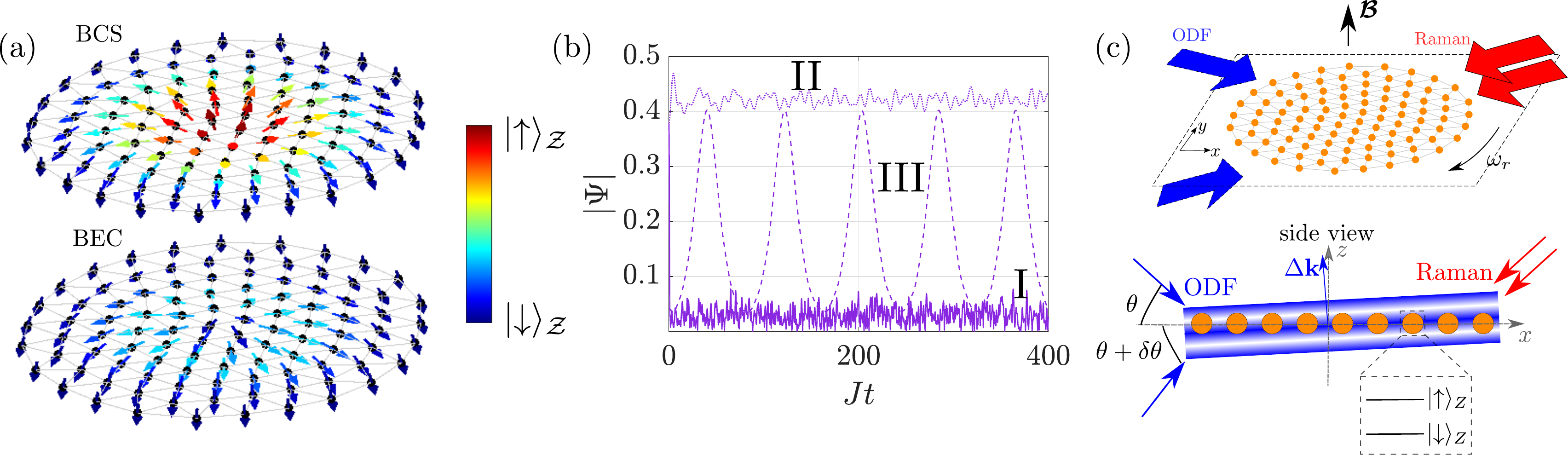}
    \caption{\textbf{Probing dynamical phases of $p+ip$ superfluids using ion crystals in Penning traps.} (a) The fermionic model is mapped on to spins encoded in the internal states of the ions, where spin up (down) represents the presence (absence) of a Cooper pair (Anderson pseudospin mapping). Here we show representative spin textures that can be engineered for the topologically nontrivial BCS and trivial BEC phases. (b) The dynamical phases are classified according to the long time behavior of the magnitude of an order parameter--- Phase I: $\abs{\Psi(t)}\rightarrow 0$; Phase II: $\abs{\Psi(t)}\rightarrow$ non-zero constant; Phase III: $\abs{\Psi(t)}$ displays persistent oscillations. (c) Schematic of our experimental proposal. State initialization, $p$-wave interactions and readout are all achieved using appropriate parameters for a pair of optical dipole force (ODF) lasers and a pair of copropagating Raman lasers. In contrast to prior implementations, the ODF difference wavevector $\delta \mbf{k}$ has both an out-of-plane and in-plane component (see side view). The result is a tilted traveling wave lattice that crosses the crystal plane slightly obliquely and thus couples the ions' electronic degrees of freedom, the out-of-plane center-of-mass mode and the in-plane crystal rotation. The spin-space directions $\mcZ$ in panel (a) and $Z$ in panel (c) are related by a rotation, as discussed in Sec.~\ref{sec:implement}.}
    \label{fig:fig1}
\end{figure*}


\section{Introduction}

The observation and classification of dynamical behaviors in quantum many-body systems constitute  a core  milestone in    quantum science. 
One  fascinating and promising paradigm comprises the  dynamical  phases    predicted   to   emerge   from   quenches  of  superconductors and   superfluids~\cite{levitov2006PRL,yuzbashyan2015PRA,lewisswan2021PRL}, systems  that feature Cooper pairing of electrons or neutral fermionic atoms. In particular,  topological $p +ip$ Bardeen–Cooper–Schrieffer (BCS) superconductors (in charged electrons) or  superfluids (in neutral atoms)--- systems that feature non-trivial topological properties~\cite{foster2013PRB} and  gapless, chiral edge states that circulate around the boundary--- are especially exciting given their potential use  for topological quantum computation.

Despite intensive theoretical efforts, $p+ip$ superfluids have eluded experimental observation, with the only confirmed realization
being the A-phase of ${}^3 {\rm
He}$, which ironically is one of the oldest-known superfluids but is also hard to control and manipulate. The realization of $p+ip$ superfluids in ultracold fermionic quantum gases, which are currently the leading platform for quantum simulation of strongly correlated matter, has also proved to be difficult. The reason is that, in spite of  all the attractive features of ultracold quantum gases, the control and manipulation of $p$-wave interactions in these systems  has remained a challenge since $p$-wave interactions  are weak under standard conditions and require Feshbach resonances to enhance them. The latter unfortunately introduce strong three-body processes which make the gas unstable and destroy the desired  pairing processes \cite{Regal:2003go,regalp2003,Schunck:2005cf,Gaebler:2007,Gunter2005}, although schemes circumventing this problem have been proposed~\cite{han2009PRL,cooper2009PRL,fedorov2016SciRep}. Theory proposals have also suggested the observation of topological superfluids  by suddenly bringing weakly interacting atoms close to a Feshbach resonance~\cite{foster2014PRL}, but to date experimental efforts remain unsuccessful.

In the present work, we propose a pathway towards the observation of non-equilibrium dynamical phases of topological $p+ip$  superfluids  by using  a 2D crystal  of ions in a Penning trap. This platform offers a high degree of control and flexibility in state initialization, interaction control and readout, that have been previously leveraged for the sensing of weak electric fields and for the simulation of quantum magnets~\cite{gilmore2021Science,garttner2017NatPhys}. In this system, we propose to  encode a spin-1/2 degree of freedom in two electronic states of the ions, which, via the Anderson pseudospin mapping~\cite{anderson1958PR}, are used  to simulate the presence or the absence of a Cooper pair. 

Our proposal takes advantage of the fact that the ion crystal in a Penning trap is rotating in the lab frame~\cite{shankar2020PRA}. This feature has never before been exploited in the context of quantum simulation and, in fact,  it is often viewed  as an impediment, e.g., to perform single site addressing. We show that by tuning the orientation and parameters of the laser beams that are typically used to couple the electronic and motional degrees of freedom of the crystal, we can engineer controllable effective interactions that simulate the  Hamiltonian of a $p+ip$ superfluid. Tuning the laser parameters also allows us to 
i) prepare initial states that resemble the low energy conditions of a $p$-wave superfluid,  ii)  control the relative strength between the kinetic energy and pairing interaction terms in order to observe the three different mean-field dynamical phases predicted to exist in $p+ip$ superfluids~\cite{foster2013PRB} and 
iii) measure a superconducting order parameter for classifying the dynamical phases.  Moreover, since  state-of-the-art ion crystals are not in the thermodynamic limit  but are instead limited to $\lesssim$ 500 ions, they naturally open a path to explore modifications to the non-equilibrium dynamics arising from beyond-mean-field effects. 
 
A key appeal  of $p+ip$ superfluids compared to ordinary superfluids is the possibility of featuring states with  nontrivial topological order. In an ordinary superfluid or superconductor, the BCS  and the BEC regimes--- which respectively  favor   weakly bound Cooper pairs and a Bose-Einstein Condensate of tightly bound molecules made of two fermions--- 
are continuously connected and are only distinguished by the strength of the pairing.  In contrast, the two regimes exhibit different topological behaviors in 2D $p+ ip$  superfluids, with a genuine quantum phase transition separating the 
topologically nontrivial BCS phase from the topologically trivial BEC phase in the equilibrium situation. This feature extends into the non-equilibrium regime, where the dynamical phases exhibit a dynamical topological quantum phase transition~\cite{foster2013PRB}.
Here, we  show  how to engineer both topologically trivial and nontrivial dynamical phases in our system and demonstrate how their topological character can be distinguished by inferring an appropriate winding number and additionally confirmed by  measurements of the effective  Cooper pair distribution function.
 
\section{2D $p+ip$ superfluids} 

The  Hamiltonian that governs the low energy  sector  of $p+ip$ superfluids is given (setting $\hbar=1$) by 
\begin{eqnarray}
\hat{H} = \sum_{\mathbf{p}} \frac{p^2}{2m} \hat{c}_{\mathbf{p}}^\dagger \hat{c}_{\mathbf{p}}
-\frac{\mathcal{G}}{2m}\sum_{\mathbf{p},\mathbf{q}} \mathbf{p}\cdot \mathbf{q} \hat{c}_{\mathbf{p}}^\dagger \hat{c}_{-\mathbf{p}}^\dagger
\hat{c}_{-\mathbf{q}} \hat{c}_{\mathbf{q}}.
\label{eqn:pwave_fermionic}
\end{eqnarray}
Here, $m$ is the electron mass and $\hat{c}_{\mathbf{p}}^\dagger, \hat{c}_{\mathbf{p}}$ are fermionic creation and annihilation operators for a fermion with momentum $\mathbf{p}$. The first term describes the single particle dispersion and the second term the attractive ($\mathcal{G}>0$, dimensionless) $p$-wave interactions that lead to the formation of Cooper pairs. For the case of 2D $p+ip$ superconductors, the momentum $\mathbf{p}\equiv p_x \hat{\mathbf{e}}_x + p_y \hat{\mathbf{e}}_y$ is restricted to the $x-y$ plane. This Hamiltonian assumes that Cooper pairs are only created and destroyed with zero center-of-mass momentum and neglects pair-breaking processes.

Under these conditions,  the low-energy physics can be mapped on to the dynamics of a collection of interacting spin-$1/2$ systems via the Anderson pseudospin mapping~\cite{anderson1958PR} that introduces spin-$1/2$ operators at each momentum $\mathbf{p}$:
\begin{eqnarray}
&&2\hS_{\mathbf{p}}^Z = \hat{c}_{\mathbf{p}}^\dagger \hat{c}_{\mathbf{p}} + \hat{c}_{-\mathbf{p}}^\dagger \hat{c}_{-\mathbf{p}}
-1, \nonumber\\
&&\hS_{\mathbf{p}}^+ = \hat{c}_{\mathbf{p}}^\dagger \hat{c}_{-\mathbf{p}}^\dagger, \;
\hS_{\mathbf{p}}^- = \hat{c}_{-\mathbf{p}} \hat{c}_{\mathbf{p}}.
\end{eqnarray}
Here,  the presence or absence of a Cooper pair at momentum $\mathbf{p}$ corresponds to  the eigenstates $\ket{\uparrow}_{\mathbf{p}},\ket{\downarrow}_{\mathbf{p}}$ of $\hS_{\mathbf{p}}^Z$ with eigenvalues $\pm 1/2$ respectively, and  the raising and lowering operators $\hS_{\mathbf{p}}^\pm$ describe the creation and annihilation of this Cooper pair.
 In terms of the Anderson pseudospin operators, Hamiltonian~(\ref{eqn:pwave_fermionic}) can be expressed as~\footnote{When converting Eq.~(\ref{eqn:pwave_fermionic}) to Eq.~(\ref{eqn:pwave_anderson}), the summation in the spin model is restricted to $\mathbf{p},\mathbf{q}$ in one half-plane, say $p_y,q_y \geq 0$, so that all the spin operators are independent~\cite{foster2013PRB}. However, extending the summation to the full plane only increases the number of spins and does not change the physics, and hence we relax the half-plane restriction in writing Eq.~(\ref{eqn:pwave_anderson}).} 
\begin{eqnarray}
\hat{H} = {\sum_{\mathbf{p}}} \frac{p^2}{m} \hS_{\mathbf{p}}^Z
-\frac{2\mathcal{G}}{m} {\sum_{\mathbf{p},\mathbf{q}}} \mathbf{p}\cdot \mathbf{q} \hS_{\mathbf{p}}^+ \hS_{\mathbf{q}}^-.
\label{eqn:pwave_anderson}
\end{eqnarray}


The ground state of the spin model~(\ref{eqn:pwave_anderson}) possesses the property that the spin orientation is correlated with the azimuthal angle $\phi_{\mathbf{p}}$ in momentum space, giving rise to chiral spin textures as depicted in Fig.~\ref{fig:fig1}(a). A winding number $Q$ can be ascribed to the spin texture, based on which the state can be classified as belonging to a topologically non-trivial BCS phase ($Q=1$) or a topologically trivial BEC phase ($Q=0$) phase. Physically, assuming spins at large momenta are always  held fixed in $\ket{\downarrow}$, the spin texture is topologically non-trivial if the central spin at $\mbf{p}=0$ is in $\ket{\uparrow}$ whereas it is trivial if this spin is in $\ket{\downarrow}$. 

The quench dynamics of $p+ip$ superconductors were theoretically studied~\cite{foster2013PRB} by considering a chiral variant of the spin model~(\ref{eqn:pwave_anderson}) given by 
\begin{eqnarray}
\hat{H} = {\sum_{\mathbf{p}}} \frac{p^2}{m} \hS_{\mathbf{p}}^Z
-\frac{\mathcal{G}}{m} {\sum_{\mathbf{p},\mathbf{q}}} p q e^{-i\left(\phi_{\mathbf{p}}-\phi_{\mathbf{q}}\right)}\hS_{\mathbf{p}}^+ \hS_{\mathbf{q}}^-,
\label{eqn:pwave_chiral}
\end{eqnarray}
where $p$ and $\phi_{\mathbf{p}}$ are the magnitude and azimuthal angle for the 2D momentum $\mathbf{p}$. This Hamiltonian  breaks time-reversal symmetry explicitly, and
preferentially selects $p_x - i p_y$ over $p_x + i p_y$ pairing, which  are degenerate in the time-reversal-invariant Hamiltonian~(\ref{eqn:pwave_fermionic}). Nevertheless both Hamiltonians possess the same $p+ip$ ground state and dynamical phases in the thermodynamic limit.  Therefore, in this work, we focus on the quantum simulation of the chiral model~(\ref{eqn:pwave_chiral}).

In the thermodynamic limit, where mean-field theory is exact, the dynamics can be pictured as  each spin precessing about a local magnetic field:
\begin{eqnarray}
\frac{d}{dt}\ev{\hat{\boldsymbol{s}}_\mathbf{p}} =  \ev{\hat{\boldsymbol{s}}_\mathbf{p}} \times \mathbf{B}_\mathbf{p}.
\end{eqnarray}
Here, $\ev{\hat{\boldsymbol{s}}_\mathbf{p}}$ is the expectation value of the spin vector at momentum $\mathbf{p}$ and $\mathbf{B}_\mathbf{p}$ is the local magnetic field with components 
\begin{eqnarray}
B_\mathbf{p}^X &=& -p\cos(\phi_\mathbf{p}) \mathrm{Re}\left[\Psi\right]- p\sin(\phi_\mathbf{p}) \mathrm{Im}\left[\Psi\right], \nonumber\\
B_\mathbf{p}^Y &=& p\cos(\phi_\mathbf{p}) \mathrm{Im}\left[\Psi\right]- p\sin(\phi_\mathbf{p}) \mathrm{Re}\left[\Psi\right], \nonumber\\
B_\mathbf{p}^Z &=& -\frac{p^2}{m},
\label{eqn:b_local}
\end{eqnarray} 
written in terms of an order  parameter $\Psi$ given by  
\begin{eqnarray}
\Psi(t) = -\frac{2\mathcal{G}}{m}{\sum_{\mathbf{p}}} p e^{i\phi_{\mathbf{p}}}\ev{\hS_{\mathbf{p}}^-(t)}.
\end{eqnarray}

Mean-field theory predicts the emergence of three dynamical phases when the system is initialized in its ground state and the pairing strength $\mathcal{G}$ is quenched. They  have been  classified according to the long-time behavior of $\abs{\Psi(t)}$ as illustrated in 
Fig.~\ref{fig:fig1}(b), where we plot a dimensionless and normalized version of the order parameter, [see Eq.~(\ref{eqn:ord_trap})]. In Phase I, the single-particle kinetic energy term ($B_\mathbf{p}^Z$)  dominates and $\abs{\Psi(t)}\rightarrow 0$. In phases II and III  interactions instead  stabilize a finite order parameter. In Phase II, $\abs{\Psi(t)}$ tends to a non-zero constant value, while  in Phase III, also known as a self-generated Floquet phase~\cite{Barankov2006}, $\abs{\Psi(t)}$  features persistent oscillations. 

The topological properties of the dynamical phases are best understood in terms of a second winding number $W$, which can take on a non-trivial value of $1$ in phases II and III. Although this quantity is formally defined in terms of retarded single-particle Green functions~\cite{foster2013PRB}, it can be physically interpreted in Phase II as the winding of the magnetic field texture in an appropriate rotating frame. In the thermodynamic limit, mean-field theory predicts that the long-time order parameter in phase II can be written as 
\begin{eqnarray}
\Psi(t) =\Psi_{\infty} e^{-2i\mu_{\infty} t},
\label{eqn:phase2_psi}
\end{eqnarray}
where $\Psi_{\infty}$ is the magnitude in the limit $t\rightarrow \infty$ and $\mu_{\infty}$ is a dynamical chemical potential. In a frame rotating at $2\mu_{\infty}$, the spins precess under a static effective magnetic field $\overline{\mathbf{B}}_\mathbf{p}$ whose  texture can be analogous to the spin texture in a $p+ip$ ground state. The winding number $W$ is computed as   
\begin{eqnarray}
W = \frac{1}{4\pi}\int d p_x d p_y \hat{\overline{\mathbf{B}}}_{\mathbf{p}} \cdot \left(\frac{d \hat{\overline{\mathbf{B}}}_{\mathbf{p}} }{d p_x} \times \frac{d \hat{\overline{\mathbf{B}}}_{\mathbf{p}} }{d p_y} \right),
\label{eqn:winding}
\end{eqnarray}
where $\hat{\overline{\mathbf{B}}}_\mathbf{p}$ denotes the corresponding unit vector in the rotating frame. In particular, the $Z$-component of $\overline{\mathbf{B}}_\mathbf{p}$ is given by $\overline{B}_\mathbf{p}^Z = B_\mathbf{p}^Z + 2\mu_\infty\hat{\mathbf{e}}_Z$. While the spins at large momenta $p\rightarrow \infty$ experience a field $\overline{\mathbf{B}}_\mathbf{p} \approx -(p^2/m) \hat{\mathbf{e}}_Z$ that points down, the central spin at $\mathbf{p}=\boldsymbol{0}$ is isolated from the other spins and experiences an effective magnetic field $\overline{\mathbf{B}}_{\boldsymbol{0}} = 2\mu_\infty \hat{\mathbf{e}}_Z$. Therefore, the magnetic field texture is BCS-like and topologically nontrivial ($W=1$) for $\mu_\infty>0$ ($\overline{\mathbf{B}}_{\boldsymbol{0}}$ pointing up) while it is BEC-like and topologically trivial ($W=0$) for $\mu_\infty<0$ ($\overline{\mathbf{B}}_{\boldsymbol{0}}$ pointing down).

\section{\label{sec:implement}Implementation with Penning traps}

We now discuss how the spin model~(\ref{eqn:pwave_anderson}) can be simulated with ion crystals in a Penning trap, where the pseudospin-$1/2$ system is  encoded in two long-lived hyperfine states of each trapped ion. In this trap, ions self-organize into a planar crystal with an approximate triangular lattice structure under the influence of static trapping fields~\cite{wang2013PRA}.  An electric quadrupole field $\boldsymbol{\mathcal{E}}$ accomplishes axial trapping and confines the ions to a single plane. The addition of a strong axial magnetic field $\boldsymbol{\mathcal{B}}$ leads to an $\boldsymbol{\mathcal{E}} \times \boldsymbol{\mathcal{B}}$ drift of the ions in this plane. This rotation provides radial confinement and the corresponding rotation frequency $\omega_r$ can be precisely controlled by additional electrodes. The out-of-plane motion of a crystal of $N$ ions is  described using $N$ normal modes of vibration, called the drumhead modes. The highest frequency drumhead mode is the center-of-mass (c.m.) mode, which is well separated from the rest of the modes and hence can be well resolved~\cite{shankar2020PRA}. 

In our modeling, the drumhead c.m. mode is treated quantum mechanically and described by bosonic creation and annihilation operators $\hat{a}_1^\dag,\hat{a}_1$. On the other hand, the planar motion is dominated by the crystal rotation, and is hence treated classically with the $x_j$ and $y_j$ coordinates of ion $j$ undergoing uniform circular motion at radius $r_j$ from the trap center, with frequency $\omega_r$ and azimuthal phase offset $\phi_j$.

As a first step to realize Hamiltonian~(\ref{eqn:pwave_chiral}), we engineer a Jaynes-Cummings type interaction between each spin and the drumhead c.m. mode, with the coupling depending on the planar position of the ion as viewed in the crystal rotating frame. As we explain shortly, the Hamiltonian we engineer is given by 
\begin{eqnarray}
\hat{H}_\text{2ch} = &&\sum_{j=1}^N B_1 \td{r}_j^2\hSS_j^{{\mcZ}} + \delta_1 \hat{a}_1^\dagger\hat{a}_1 \nonumber\\
&&-\sum_{j=1}^N \frac{G}{i\sqrt{N}}\td{r}_j \left(\hSS_j^- \hat{a}_1^\dagger e^{i\phi_j}  - \hat{a}_1 \hSS_j^+ e^{-i\phi_j}
\right).
\label{eqn:ham_2ch}
\end{eqnarray}
This Hamiltonian is written in a rotated spin space $\mcZ\equiv -X$, $\mcX\equiv Z$, $\mcY\equiv Y$, and $\hSS_j$ denotes spin operators in this rotated space. Here, $\td{r}_j=r_j/R$ is a normalized radial coordinate, $B_1$ is a frequency controlling the dispersion of the spins, $\delta_1$ is an effective detuning of the c.m. mode from the spins and $G$ is a frequency controlling the spin-mode coupling strength. In particular, the amplitude and phase of the coupling of spin $j$ to the c.m. mode respectively depend on $\td{r}_j$ and $\phi_j$. 

Equation~(\ref{eqn:ham_2ch}) describes the Hamiltonian for the so-called two-channel model of a $p$-wave superconductor. The c.m. mode plays the role of the bosonic molecular channel, while each ion encodes an Anderson pseudospin in its electronic states. Here, spin up (down) indicates the presence (absence) of a Cooper pair. While the Anderson pseudospins live in a lattice in momentum space where the coordinates are $(p_x,p_y)$, the role of momentum is instead played here by the position $(x_j,y_j)$ of each ion $j$ in the crystal plane. `Momentum'-dependent rates appear in the single-particle and interaction terms through the radius $r_j$ and the phase factors $e^{\pm i\phi_j}$.

Subsequently, an effective spin model can be derived in the situation when $\delta_1\gg G,B_1$. Using effective Hamiltonian theory~\cite{james2007CJP} and assuming that the c.m. mode is in the motional ground state, we obtain the one-channel model given by
\begin{eqnarray}
\hat{H}_{\text{1ch}} = K\sum_{j}\td{r}_j^2 \hSS_j^{{\mcZ}} - \frac{J}{N}\sum_{j \neq k} \td{r}_j \td{r}_k \hSS_j^+ \hSS_k^- e^{-i(\phi_j-\phi_k)},
\label{eqn:ham_1ch}
\end{eqnarray}
where $J=G^2/\delta_1$ and $K=B_1-J/N$. Equation~(\ref{eqn:ham_1ch}) is essentially the one-channel $p$-wave Hamiltonian~(\ref{eqn:pwave_chiral}) that we wish to simulate.

We now briefly outline how Hamiltonian~(\ref{eqn:ham_2ch}) can be engineered while presenting the detailed derivation in Appendix~\ref{appsec:derivation}. Coupling between the spins and the crystal motion is enabled by the application of an optical dipole force (ODF) that gives rise to spatially dependent AC Stark shifts on the spin states~\cite{britton2012nat}. The ODF is generated by two traveling-wave lasers with difference wavevector $\Delta\mbf{k}$ and beatnote frequency $\mu_r$. In typical applications, only the axial motion is coupled to the spin and hence $\Delta\mbf{k}\parallel \hat{z}$~\cite{gilmore2021Science,garttner2017NatPhys}. However, in this work, we consider the $\Delta\mbf{k}$ to have non-zero components both along the $\hat{z}$ and $\hat{x}$ directions~[Fig.~\ref{fig:fig1}(c)]. The result is a spatially varying AC Stark shift that depends on both the in-plane and out-of-plane motions of the ions, thereby coupling the spins to the motion along both the directions. 

A second ingredient in our proposal consists of a pair of co-propagating Raman lasers that drives spin flips without coupling to the motion. We assume that the two Raman lasers have an identical but tunable beam waist $w$, leading to an effective two-photon Rabi frequency that is radially varying as $B(r) = B_0 e^{-r^2/w^2}$ and corresponding Hamiltonian $\hat{H}_{\text{Raman},j} = B(r_j)\hS_j^X$. For $w\gg R$, where $R$ is the crystal radius, we can approximate $B(r)\approx B_0 - B_0 r^2/w^2$.

The role of the Raman drive is twofold and becomes apparent in the rotated spin space (see Appendix~\ref{appsec:derivation}). First, the spatially homogeneous drive with strength $B_0$ serves to break the symmetry between a Jaynes-Cummings and an anti-Jaynes-Cummings type interaction of the spins and the drumhead c.m. mode that arise due to the ODF. The Jaynes-Cummings term can then be selectively brought near resonance by an appropriate choice of the ODF beatnote frequency $\mu_r$. Second, the beam waist $w$ serves as a control knob for tuning the single particle dispersion, i.e. $B_1=B_0R^2/w^2$. We note that the Raman beams can be replaced with a microwave drive that limits the tunability of $B_1$ but allows for a simpler implementation and reduced decoherence (see Appendix~\ref{appsec:params}). 

We present potential experimental parameters for realizing our proposal in Appendix~\ref{appsec:params} and study the adverse impact of off-resonant terms in Appendices~\ref{appsec:rwa} and \ref{appsec:unwanted}. Our study suggests that it is possible to operate in parameter regimes where the off-resonant terms have only a small effect. Although our study of off-resonant terms is extensive, our analysis of their impact is not exhaustive because of the sheer number of such terms. 
Their impact and the parameter regimes where they are negligible could potentially be explored directly on the quantum simulator. In addition, for typical operating conditions, we estimate that decoherence from off-resonant light scattering may limit the simulation time. However, we note that the relative strength of coherent interaction to decoherence can be increased, for instance, by the choice of ion species and transition, by enhancing coherent coupling via parametric amplification~\cite{ge2019PRL} or by working at a different operating point for the optical dipole force.

\subsection{Initialization and readout}

To observe dynamical phases generated by Hamiltonian~(\ref{eqn:ham_1ch}), the spins must be initialized in a state possessing a chiral spin texture with a nonzero order parameter, such as the ones shown in Fig.~\ref{fig:fig1}(a). In the case of the trapped ion crystal, it is convenient to work with a normalized order parameter $\Psi$ defined as
\begin{eqnarray}
\Psi(t) = \frac{2}{N}\sum_{j=1}^N \td{r}_j e^{i\phi_j}\ev{\hSS_j^-(t)}.
\label{eqn:ord_trap}
\end{eqnarray}
For engineering interactions, the ODF beanote frequency $\mu_r$ was tuned to couple the spins, the drumhead c.m. mode and the planar rotation. For preparing chiral initial states, we instead tune $\mu_r$ to only couple the spins to the planar rotation without involving the drumhead c.m. mode. By additionally tuning the beam waist $w_\mathrm{ODF}$ of the ODF lasers, both BCS and BEC-like initial spin textures can be prepared. The initialization Hamiltonian is of the form 
\begin{eqnarray}
\hat{H}_\text{init}=\frac{\Omega_0}{2}\sum_j e^{-r_j^2/{w_\mathrm{ODF}^2}} \td{r}_j\left(\hSS_j^+ e^{-i\phi_j} + \hSS_j^- e^{i\phi_j} \right),
\label{eqn:h_init}
\end{eqnarray}
where $\Omega_0$ is an effective drive strength. This Hamiltonian drives single-spin rotations where the axis of rotation for ion $j$ depends on the azimuthal angle $\phi_j$ in the rotating frame. Starting with all spins initialized in $\ket{\uparrow}_{{\mcZ}}$, setting $w_\mathrm{ODF}\gg R$, and using a pulse area $\Omega_0 T=\pi$ results in a BCS-like spin texture~[e.g. Fig.~\ref{fig:fig1}(a)], that can be used to observe phases I and II. Here, we have exploited the fact that the magnitude of the Rabi frequency increases with the radius so that the central spin is unaffected while the outermost spins are rotated to $\ket{\downarrow}_{\mcZ}$. A BEC-like spin texture~[e.g. Fig.~\ref{fig:fig1}(a)] can be engineered by setting $w_\mathrm{ODF}<R$ and starting with all spins in $\ket{\downarrow}_{{\mcZ}}$. In this way, ions at the center, where   $\td{r}_j\sim 0$, are unaffected  whereas ions at the crystal boundary are also left unchanged since the beam intensity tapers to zero. On the other hand, ions in the intermediate region experience some degree of spin rotation and thereby give rise to a BEC-like texture. For observing phase III dynamics, a BCS-like initial state with a sharp domain wall between $\ket{\downarrow}_{{\mcZ}}$ and $\ket{\uparrow}_{{\mcZ}}$ spins and a small value of $\abs{\Psi(0)}$ is suitable~[e.g. top panel of Fig.~\ref{fig:fig2}(c)]~\cite{foster2013PRB}. Starting with all spins in $\ket{\downarrow}_Z$, a domain wall can be realized by using an optical pumping beam to selectively address ions in the central region and prepare them in $\ket{\uparrow}_Z$. Next, a global $\pi/2$ pulse rotates $Z\rightarrow \mcZ$ so that ions in the central region and those outside are respectively prepared in $\ket{\uparrow}_{\mcZ}$ and $\ket{\downarrow}_{\mcZ}$. A small initial $\abs{\Psi(0)}$ can then be induced by a short-time application of  Hamiltonian~(\ref{eqn:h_init}). A detailed description of state initialization is presented in Appendix~(\ref{appsec:init}).

Measurement of the real and imaginary parts of the order parameter is also enabled by the Hamiltonian~(\ref{eqn:h_init}). To demonstrate this, we first introduce site-dependent orthogonal axes, 
\begin{eqnarray}
\hat{\mathbf{e}}_{{\mcX_j'}} = \sin\phi_j \hat{\mathbf{e}}_{{\mcX}} - \cos\phi_j \hat{\mathbf{e}}_{{\mcY}},\nonumber\\
\hat{\mathbf{e}}_{{\mcY_j'}} = \cos\phi_j \hat{\mathbf{e}}_{{\mcX}} + \sin\phi_j \hat{\mathbf{e}}_{{\mcY}},
\label{eqn:local_axes}
\end{eqnarray}
such that $\hat{\mathbf{e}}_{{\mcX_j'}}\times \hat{\mathbf{e}}_{{\mcY_j'}} = \hat{\mathbf{e}}_{{\mcZ}}$. In terms of these axes, we can write 
\begin{eqnarray}
\mathrm{Re}\left[\Psi\right] = \frac{2}{N}\sum_j \td{r}_j\ev{\hSS_j^{{\mcY_j'}}},\;
\mathrm{Im}\left[\Psi\right] = \frac{2}{N}\sum_j \td{r}_j\ev{\hSS_j^{{\mcX_j'}}},\nonumber\\
\end{eqnarray}
and with $w_\mathrm{ODF}\gg R$, Hamiltonian~(\ref{eqn:h_init}) can be expressed as
\begin{eqnarray}
\hat{H}_\mathrm{init} = \Omega_0\sum_j \td{r}_j \hSS_j^{{\mcY_j'}}.
\label{eqn:h_init_compact}
\end{eqnarray}
After running the quantum simulation up to some time $T$, we  evolve the system  under (\ref{eqn:h_init_compact}) for a further time $t$. This  leads to 
\begin{eqnarray}
\ev{\hSS_j^{{\mcZ}}(T+t)} &=& \ev{\hSS_j^{{\mcZ}}(T)} \cos\left(\Omega_0\td{r}_j t\right) - \langle\hSS_j^{{\mcX_j'}}(T)\rangle \sin\left(\Omega_0\td{r}_j t\right) \nonumber\\
&\approx& \ev{\hSS_j^{{\mcZ}}(T)} - \left(\Omega_0\td{r}_j\langle\hSS_j^{{\mcX_j'}}(T)\rangle\right)t,
\end{eqnarray}
where the approximation holds true for short rotation times. Summing over all the ions, we get 
\begin{eqnarray}
\ev{\hat{\mathfrak{J}}^{{\mcZ}}(T+t)} - \ev{\hat{\mathfrak{J}}^{{\mcZ}}(T)} \approx -\frac{N\Omega_0}{2}\mathrm{Im}\left[\Psi(T)\right]t,
\end{eqnarray}
where $\hat{\mathfrak{J}}^{{\mcZ}} = \sum_j \hSS_j^{{\mcZ}}$. The imaginary part of the order parameter determines the rate of change of $\ev{\hat{\mathfrak{J}}^{{\mcZ}}(T+t)}$ as the rotation time $t$ is increased. This quantity is accessible by a global fluorescence measurement after applying a  global  $\pi/2$ pulse to rotate $\mcZ\rightarrow Z$. Furthermore, a phase offset of $\pi/2$ can be introduced in Hamiltonian~(\ref{eqn:h_init}) by shifting the phase of the ODF beatnote, which can be used to measure $\mathrm{Re}\left[{\Psi(T)}\right]$ in a similar manner. 


\section{Probing the  dynamical phases}

\begin{figure*}[tb]
    \centering
    \includegraphics[width=\textwidth]{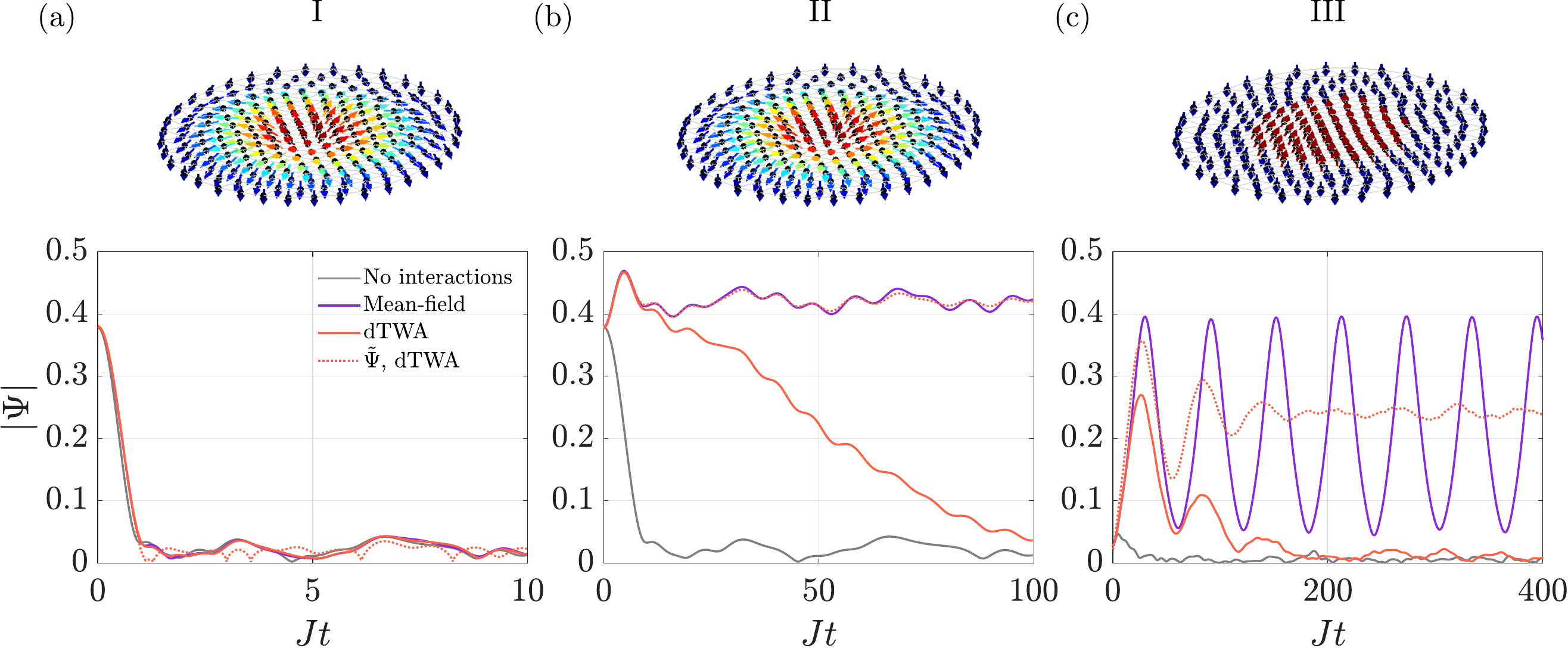}
    \caption{\textbf{Manifestation of dynamical phases in a 200 ion crystal.} Initial BCS-like spin textures (top panels) and time evolution of $\abs{\Psi(t)}$ in phases (a) I, (b) II and (c) III. In (a) $K/J=10$, while in (b) and (c) $K/J=1$. The dynamics of $\abs{\Psi(t)}$ are computed both using mean-field theory and the discrete truncated Wigner approximation (dTWA) method. The dTWA results show that the finite size of the crystal leads to a decay of $\abs{\Psi(t)}$ even in phases II and III. However, the buildup of quantum correlations in these phases is captured by a second order parameter $\td{\Psi}$ [Eq.~(\ref{eqn:psi_td})]. In all cases, the decay of $\abs{\Psi(t)}$ in the absence of interactions is plotted for reference. 
    Crystal parameters are detailed in Appendix~\ref{appsec:params} and chiral spin states are initialized according to Appendix~\ref{appsec:init}.}
    \label{fig:fig2}
\end{figure*}

Having established protocols for initializing BCS-like and BEC-like initial states, for engineering interactions, and for measuring the order parameter, we now proceed to discuss how the dynamical phases can be observed  in a crystal stored in a Penning trap. Figure~\ref{fig:fig2} shows representative examples of the three dynamical phases, which are obtained using different BCS-like initial conditions and interaction strengths, the latter characterized by the ratio $K/J$. In each case, the initial spin texture is shown in the top panel. Phases I and II use the same initial spin texture but are obtained using $K/J=10$ and $K/J=1$ respectively. On the other hand, phase III is obtained using a different initial spin texture but with the same value of $K/J=1$ as in phase II. The purple lines in  Fig.~\ref{fig:fig2} show the mean-field predictions for the time evolution of $\abs{\Psi(t)}$ in each phase. The magnitude $\abs{\Psi(t)}$ decays to $0$ in phase I, saturates to a non-zero constant in phase II and displays persistent oscillations in phase III. However, given the finite number of ions ($N=200$), we are  motivated to investigate the validity of mean-field theory in describing our system. Towards this end, we simulate the dynamics under $\hat{H}_\text{1ch}$ via the discrete truncated Wigner approximation (dTWA) method that accounts for the quantum noise of the initial state~\cite{scachenmayer2015PRX} (see Appendix~\ref{appsec:num}). The results of the dTWA simulations are shown by the red lines in Fig.~\ref{fig:fig2}. The dTWA and mean-field results agree well in Phase I where the single-particle dephasing dominates the interactions. In contrast, the dTWA results deviate significantly from the mean-field predictions in Phases II and III. In both cases, quantum noise causes the order parameter to eventually decay to zero in the long time limit. 

Importantly, the decay of the mean-field order parameter $\abs{\Psi(t)}$ in phases II and III is accompanied by the development of quantum correlations, which is captured in a more appropriate  order parameter $\td{\Psi}$ defined as 
\begin{eqnarray}
\td{\Psi} = \frac{2}{N}\abs{\sum_{j\neq k}\td{r}_j\td{r}_k\ev{\hSS_j^+\hSS_k^-}e^{-i(\phi_j-\phi_k)}}^{1/2}.
\label{eqn:psi_td}
\end{eqnarray}
We note that $\td{\Psi}$ is just a measure of the interaction part of the Hamiltonian $\hat{H}_\text{1ch}$ [Eq.~(\ref{eqn:ham_1ch})]. While $\Psi$ is the standard order parameter in superconductors, $\td{\Psi}$ could be thought of as the density of the Cooper pairs without concern to whether they are condensed. This is similar to the BEC phase of the BCS-BEC condensates, kept above the superconducting transition temperature and below the temperature of the formation of pairs~\cite{randeria2014AnnRev}. Figure~\ref{fig:fig2} shows that $\td{\Psi}$ stabilizes to a non-zero constant in phases II and III indicating the persistence of dynamical $p$-wave superfluidity in these phases. 

Even though the mean-field order parameter $\abs{\Psi(t)}$ disappears at long times, the three phases can  be distinguished  in the short time dynamics of this observable. Figure~\ref{fig:fig2} shows that  for  $Jt\lesssim 20$, the mean-field and dTWA results are in approximate agreement in all three phases. The magnitude $\abs{\Psi(t)}$ does not decay in Phase II whereas it exhibits strong amplification in Phase III.
In an experiment, the stabilization  of superfluidity by  interactions  can be sharply demonstrated by comparing the dynamics of $\abs{\Psi(t)}$ in the presence and absence of the ODF drive; in the absence of interactions, $\abs{\Psi(t)}$  decays towards zero even on short time scales (gray lines in Fig.~\ref{fig:fig2}).

\section{Inferring topology}

\begin{figure*}[!tb]
    \centering
    \includegraphics[width=0.9\textwidth]{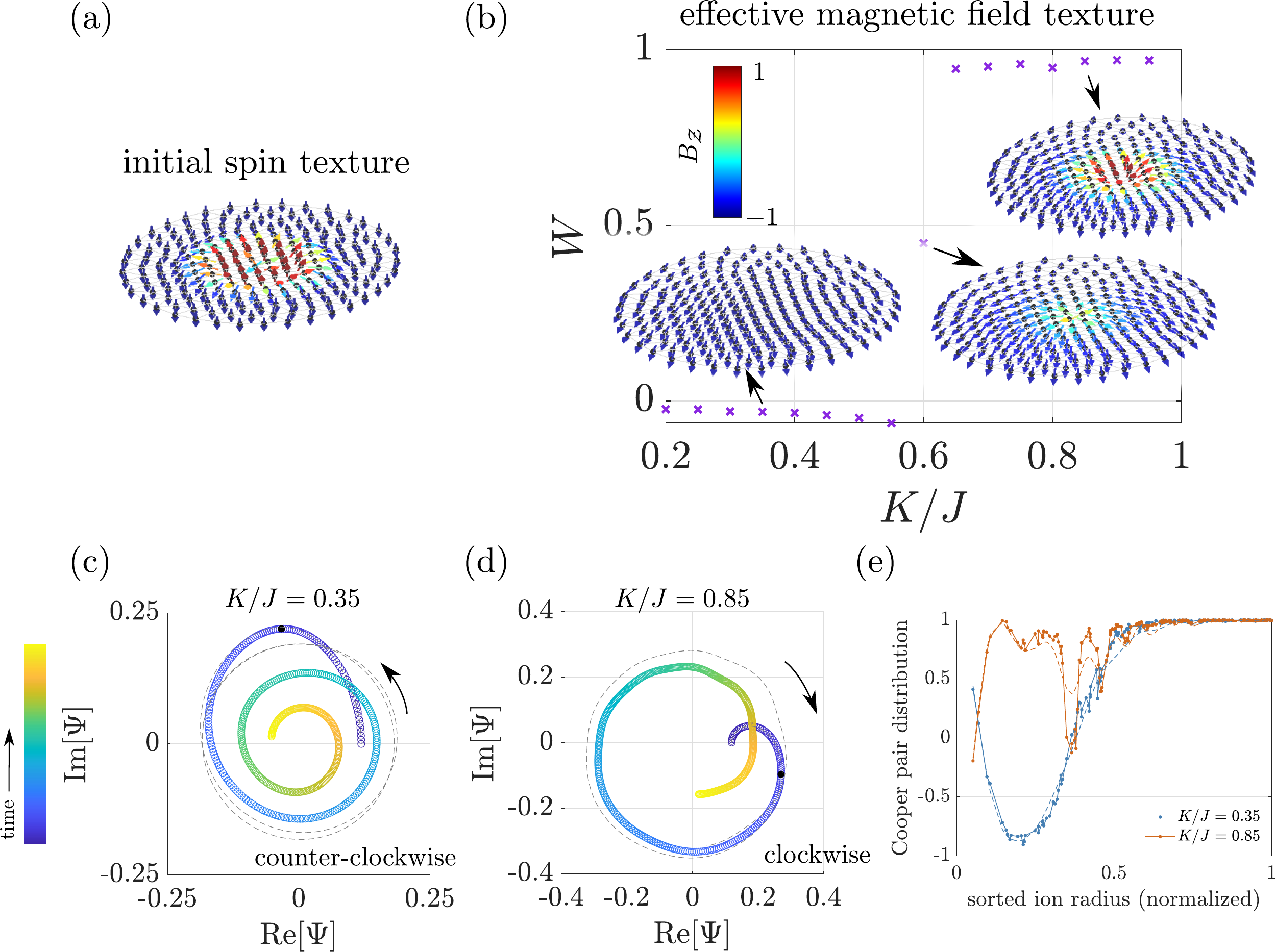}
    \caption{\textbf{Inferring topological properties of the dynamical phases.} (a) Initial BEC-like spin texture. (b) Mean-field, long-time winding number of the effective magnetic field texture as the relative strength of the single-particle ($K$) and interaction terms ($J$) are varied. In the strong interaction case ($K/J<0.6$), the texture is BEC-like and topologically trivial, while in the weak interaction case ($K/J>0.6$), it is BCS-like and topologically nontrivial. Representative field textures are shown, where each arrow now indicates the unit vector of the effective magnetic field acting at that site, and the color code indicates the normalized $B_{\mcZ}$ component. (c-d) The sense of rotation of the order parameter in the complex plane reveals the winding number of the underlying magnetic field texture. The topologically trivial (nontrivial) dynamical phase is associated with a counterclockwise (clockwise) rotation of $\Psi(t)$. While mean-field predicts a stable limit cycle (gray dashed line), dTWA calculations (color dots) show the order parameter spiralling in towards the origin at long times (plotted here until $Jt=100$). The color gradient indicates the arrow of time. 
    (e) The Cooper pair distribution function (CPDF) displays an even (odd) number of zero-crossings in the topologically trivial (non-trivial) case. These features are preserved even at times $Jt\sim 100$, when the order parameter has decayed considerably due to quantum fluctuations. For comparison, the dashed lines show the CPDF computed using mean-field theory. Crystal parameters are detailed in Appendix~\ref{appsec:params} and chiral spin states are initialized according to Appendix~\ref{appsec:init}.}
    \label{fig:fig3}
\end{figure*}

In contrast to $s$-wave superconductors, ground states and dynamical phases of $p$-wave superconductors can possess nontrivial topological properties. We now discuss how the topology of the dynamical phases can be probed in the ion simulator. For this study, we use the initial spin texture shown in Fig.~\ref{fig:fig3}(a), which is approximately BEC-like in the sense that the $\mcZ$-component of the spins first increases with radius, reaches a maximum, and then decreases with a further increase in radius. Figure~\ref{fig:fig3}(b) shows the winding number $W$ (Eq.~(\ref{eqn:winding}) computed in mean-field theory as the ratio $K/J$ is tuned for a crystal of $N=200$ ions in the Penning trap (see Appendix~\ref{appsec:winding} for details of this calculation on the discrete crystal lattice). Representative examples of the effective magnetic field texture are also shown, which demonstrate the transition from a topologically trivial BEC-like texture ($K/J < 0.6$, $W\approx 0$) to a topologically non-trivial BCS-like texture ($K/J > 0.6$, $W \approx 1$). 
Remarkably, the topologically trivial and nontrivial phases can be distinguished by measurements of the real and imaginary parts of the order parameter. From Eq.~(\ref{eqn:phase2_psi}), the sense of rotation of the order parameter in the complex plane--- clockwise ($\mu_\infty>0$) or counterclockwise ($\mu_\infty<0$)--- is a direct measurement of the sign of $\mu_\infty$, and consequently, enables us to infer the BCS-like or BEC-like nature of the effective magnetic field texture. Figures~\ref{fig:fig3}(c) and (d) show that the sense of rotation of the order parameter is different for $K/J=0.35$ and $K/J=0.85$, clearly indicating the transition from a topologically trivial to a topologically nontrivial dynamical phase as the ratio $K/J$ is tuned. While mean-field theory predicts the order parameter to precess with an approximately constant radius in the complex plane, the build-up of quantum correlations  cause the  order parameter to spiral in towards the origin at long times, consistent  with  Fig.~\ref{fig:fig2}(b).  Nevertheless,  the decay does not modify the topological nature of the dynamical phases. 

The preservation of the topology in the regime where the order parameter is decaying  can be confirmed  by measuring the  so-called Cooper pair distribution function (CPDF) $\gamma(\bf p)$~\cite{foster2013PRB}. This function characterizes  the  nonequilibrium distribution of the quasiparticle spectrum in the asymptotic steady state and  provides information  about the topology of the dynamical phases: The topology of the dynamical phase is trivial (non-trivial) if the number of zero-crossings of this function is even (odd)~\cite{foster2013PRB}. In superconductors the CPDF  can be measured via time-resolved ARPES \cite{Schwarz2020}. In the spin model, $\gamma(\bf p)$ maps to $\gamma(\mathbf{r}_j)$,  where $\mathbf{r}_j$ is the position of ion $j$, and  corresponds to the  projection of the local spin vector $\ev{{\boldsymbol{\hSS}}_j}$
onto   the local effective magnetic field $\overline{\mathbf{B}}_j$:
\begin{eqnarray}
\gamma(\mathbf{r}_j) = \frac{\ev{\overline{\boldsymbol{\hSS}}_j}\cdot \overline{\mathbf{B}}_j}{\sqrt{\ev{\overline{\boldsymbol{\hSS}}_j}\cdot\ev{\overline{\boldsymbol{\hSS}}_j}} \sqrt{\overline{\mathbf{B}}_j\cdot\overline{\mathbf{B}}_j}}. 
\end{eqnarray}
Here, the overbar denotes that the spin and effective magnetic field are measured in a frame rotating at $2\mu_\infty$ and the denominator ensures that the quantity is the direction cosine of the spin vector with respect to the local magnetic field. Figure~\ref{fig:fig3}(e) shows the CPDF as a function of ion radius from the trap center for $K/J=0.35$ and $K/J=0.85$.  The solid curves in Fig.~\ref{fig:fig3}(e) have been computed accounting for quantum fluctuations and by running the simulation up to a time $Jt=100$, for which the decay of the order parameter is clearly visible in Fig.~\ref{fig:fig3}(c-d). We find that the number of zero-crossings of the CPDF enable an inference of the topology even after significant decay of the order parameter.  We note that site-resolved measurement of the spin components is sufficient to measure the effective local magnetic field [see Eq.~(\ref{eqn:b_local})], while the value of $\mu_\infty$ can be obtained from a Fourier transform of the time series of $\Psi(t)$.

\section{Realizing a two-channel model}

\begin{figure}[!tb]
    \centering
    \includegraphics[width=0.8\columnwidth]{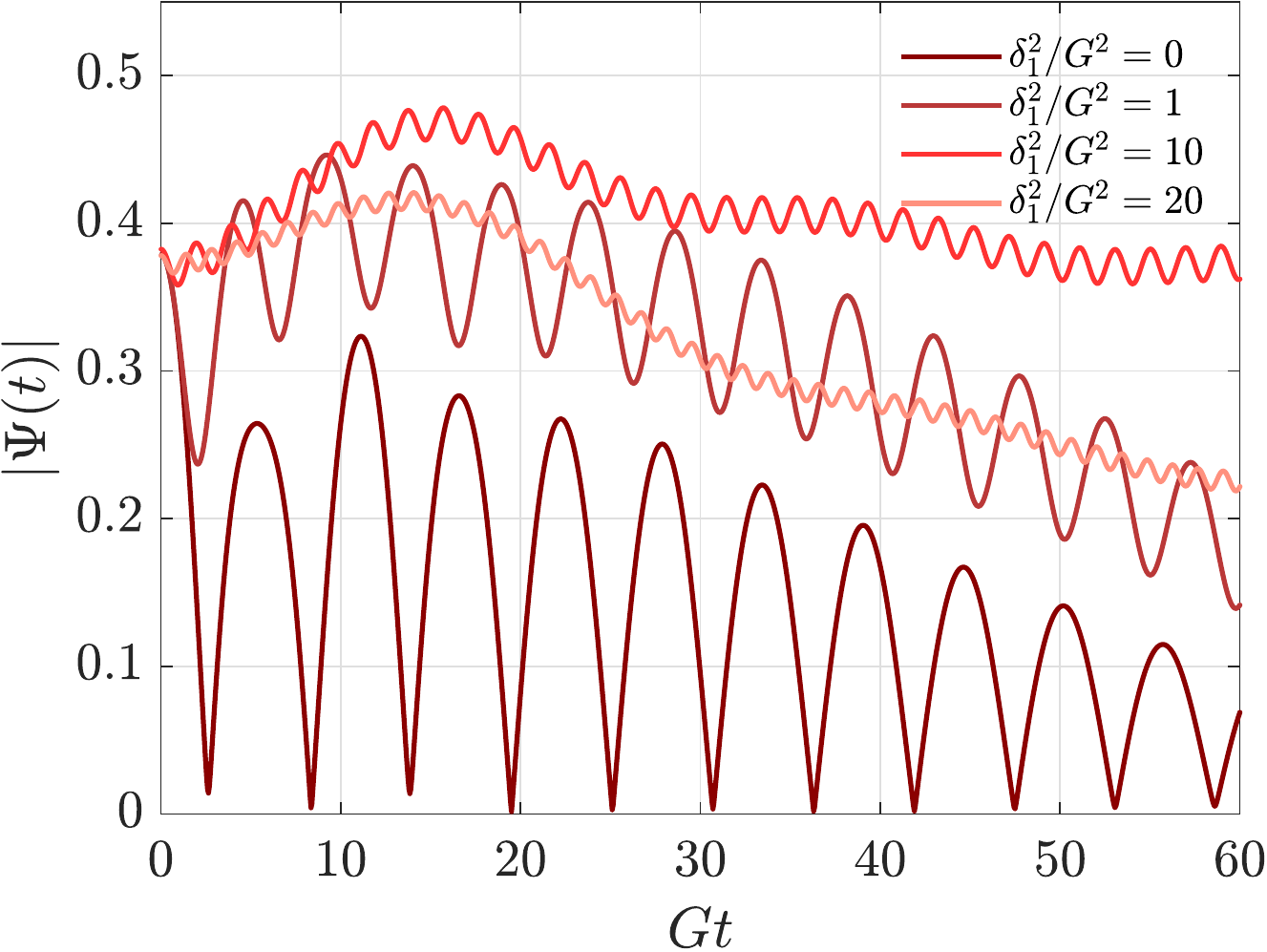}
    \caption{\textbf{Realizing the two-channel model.} The drumhead c.m. mode plays the role of the molecular channel in the two-channel $p$-wave model. As the coupling to the c.m. mode is tuned from off-resonant ($\delta_1^2/G^2\gg 1$), to near resonance ($\delta_1^2/G^2\sim 0$), its effects can be clearly observed in the time evolution of $\abs{\Psi(t)}$. Here, we fix $B_1=G/\sqrt{10}$ and vary $\delta_1$ to obtain the different curves. Crystal parameters are detailed in Appendix~\ref{appsec:params} and chiral spin states are initialized according to Appendix~\ref{appsec:init}.}
    \label{fig:fig4}
\end{figure}

So far, we have focused on a regime where the c.m. mode is coupled off-resonantly to the spins and can hence be adiabatically eliminated, giving rise to an effective one-channel model description in terms of the Anderson spins alone. The  c.m. mode plays the role of the bosonic molecular channel in the two-channel $p$-wave model. By suitably tuning the ODF difference frequency $\mu_r$, a near-resonant coupling with the c.m. mode can be engineered, thereby enabling simulation of the more general two-channel Hamiltonian~(\ref{eqn:ham_2ch}). Thus, our trapped ion simulator allows the exploration of the  distinct physical  behaviors featured by both the one-channel and two-channel models in the same experimental setting. Figure~\ref{fig:fig4} shows the time evolution of $\abs{\Psi(t)}$ as the coupling to the  c.m. mode is tuned from a far off-resonant regime ($\delta_1^2/G^2\gg 1$) to the resonant regime  ($\delta_1^2/G^2 \sim 0$). The curves are computed accounting for the initial quantum noise of the spins, which are initialized in a BCS-like state [top panel of Fig.~\ref{fig:fig2}(a)], as well as that of the c.m. mode, which is assumed to be initialized in the ground state. In the off-resonant regime, the behavior is similar to the one-channel model as the occupation of the c.m. mode remains small at all times. However, $\abs{\Psi(t)}$ exhibits pronounced oscillatory behavior in the near-resonant and resonant regimes where excitations are strongly exchanged between the spins and the c.m. mode. Interestingly, these results indicate that for fixed $B_1$, there is an optimal $\delta_1$ in order to stabilize $\abs{\Psi(t)}$ for a longer time. In addition to probing the role of the molecular channel, near-resonant coupling also enhances the interaction strength and may hence exhibit stronger signatures of the interaction dynamics in the time before sources of  decoherence such as light scattering from the ODF beams kick in (Appendix~\ref{appsec:params}).

\section{Conclusion and Outlook}

We have proposed a protocol to simulate the quench dynamics of $p+ip$ superfluids using ion crystals stored in a Penning trap. By utilizing the Anderson pseudospin mapping, we make a fermionic Hamiltonian amenable to simulation using spins encoded in the internal states of ions. In this way, we not only leverage the versatile toolbox of the trapped ion platform but we also circumvent limitations that arise with direct simulations of the fermionic model using ultracold quantum gases, where $p$-wave superfluids have been highly elusive till date. 

We demonstrated how signatures of all three dynamical phases can be observed using crystals with $\sim 200$ ions. In the thermodynamic limit, the 2D $p + ip$ model becomes integrable and the dynamics is well described by mean
field theory. However, beyond mean field effects kick in at finite number of fermions; the relatively small number of spins in the trapped ion crystal naturally opens the avenue to observe beyond mean-field effects in the quench dynamics of $p+ip$ superfluids and also facilitates the emulation of small superconducting grains~\cite{delft2001AnnDerPhys}. 

We showed how the topological character of the dynamical phases can be inferred via the sense of rotation of the order parameter in the complex plane. Since our proposal maps the fermionic momentum on to the real-space position of ions, the addition of site-resolved detection of spins in the trap can provide time and  momentum-resolved measurements of the simulated system, analogous to techniques used in solid state materials such as ARPES~\cite{liao2015PRA}. Along these lines, we showed how the Cooper pair distribution function can be measured with site-resolved detection and can be used to probe the topology of the dynamical phase. 

By tuning closer to resonance with the c.m. mode, a two-channel $p$-wave model can be realized and the nonzero occupation of a molecular channel can be additionally simulated and investigated.  It will be interesting to explore conditions where  a dynamically  active molecular channel can enhance 
superfluidity \cite{kelly2022resonant}, in a way  reminiscent  to  analogous   phenomena recently studied in solid state quantum optics\cite{Matsunaga1145,Matsunaga2013,Matsunaga2017,Mitrano2016,Mankowsky2014,Isoyama2021,Sentef2018,Jaksch2019,Curtis2019,thomas2019,Ahana2021}.
 For example, our system  can be used to study the  response when the c.m. mode (i.e. the molecular channel) is driven or even squeezed via parametric amplification~\cite{ge2019PRL,burd2021NatPhys}. Furthermore, although we have considered a regime where the other drumhead modes are off-resonant, strong spin-mode coupling can lead to weak excitations of more than a single mode, leading to the emergence of spatial inhomogeneities than can emulate the phenomenon of Cooper pair turbulence~\cite{dzero2009EPL}. 

Beyond the simulation of $p+ip$ superconductors, trapped ions could also be used to study further exotic superconducting systems such as chiral $d_{x^2- y^2} +i d_{xy}$ superfluids~\cite{fischer2014PRB}. Such systems can be simulated in the Penning trap by engineering the phase  of the ODF beams to have spatial variations, which is possible with the use of deformable mirrors~\cite{polloreno2022arXiv}.

Finally, we note that the fermionic statistics of the original $p$-wave model is not present in the corresponding spin mapping, and therefore our proposal cannot be used to produce Majorana fermions~\cite{jiang2011PRL,kraus2013PRL}. However, the protocols we have developed can be used to both produce skyrmionic spin textures as well as stabilize them using interactions, which could find applications in demonstrating skyrmion qubits for quantum computing~\cite{psaroudaki2021PRL}.  

\section*{Acknowledgements}
We thank Allison Carter for providing estimates for the decoherence rates from light scattering. We thank Diego Barberena  and Bryce Bullock  for a careful reading and comments on the manuscript. This work is supported by the European Union’s Horizon 2020 research and innovation program under Grant Agreements No. 817482 (Pasquans), by the Simons Collaboration on Ultra-Quantum Matter, which is a grant from the Simons Foundation
(651440, P. Z.), and by LASCEM via AFOSR No. 64896-PH-QC. Support is also acknowledged from the AFOSR grants FA9550-18-1-0319 and FA9550-19-1-0275, by the NSF JILA-PFC PHY-1734006, QLCI-OMA -2016244, by the U.S. Department of Energy, Office of Science, National Quantum Information Science Research Centers, Quantum Systems Accelerator, and by NIST. JJB acknowledges support from the DARPA ONISQ program and AFOSR grant FA9550-201-0019.

\balancecolsandclearpage

\begin{widetext}

\begin{appendix}

\section{\label{appsec:derivation}Derivation of effective Hamiltonians}

We show how the combination of optical dipole force (ODF) and Raman beams with suitable geometries leads to effective one-channel and two-channel models for $p$-wave interactions in ion crystals stored in a Penning trap.

\subsection{Optical dipole force with tilted wavevectors}

The Hamiltonian corresponding to only the ODF interaction is 
\begin{equation}
	\hat{H} = \sum_{j=1}^N \omega_s \hS_j^Z + \sum_{n=1}^N \omega_n \hat{a}_n^\dag \hat{a}_n + 2\sum_{j=1}^N \delta_\mathrm{AC} \sin(\Delta k_x x_j + \Delta k_z \hat{z}_j -\mu_r t) \hS_j^Z,
	\label{appeqn:ham_odf_tilted}
\end{equation}
where $\omega_s$ is the spin transition frequency, $\delta_\mathrm{AC}$ is the strength of the optical dipole force, $\Delta\mathbf{k}=\Delta k_x \hat{\mbf{e}}_x + \Delta k_z \hat{\mbf{e}}_z$ is the difference wavevector of the ODF beams and $\mu_r$ is their difference frequency. The out-of-plane motion is treated quantum mechanically and is represented by the operator $\hat{z}_j$. In terms of the drumhead modes, it can be expressed as 
\begin{eqnarray}
\Delta k_z \hat{z}_j = \sum_{n=1}^N \eta_n \mathcal{M}_{jn}\left(\hat{a}_n  + \hat{a}_n^\dagger \right),
\label{appeqn:z_expansion}
\end{eqnarray}
where $\hat{a}_n,\hat{a}_n^\dagger$ are annihilation and creation operators for mode $n$ with frequency $\omega_n$, $\eta_n=\Delta k_z\sqrt{\hbar/(2m\omega_n)}$ is the Lamb-Dicke parameter, $m$ is the ion mass, and $\mathcal{M}_{jn}$ is the displacement of ion $j$ under the influence of mode $n$. In contrast, the in-plane motion is dominated by the crystal rotation and is represented by the classical coordinate $x_j(t) = r_j \cos(\omega_r t +\phi_j)$, where $r_j$ is the distance of the ion from the trap center and $\phi_j$ is the azimuthal angle in the rotating frame. 

\subsection{Co-propagating Raman beams}

We now introduce a pair of Raman beams that drive resonant two-photon transitions between the spin states. The beams are assumed to be copropagating so that their difference wavevector approximately vanishes~\footnote{We discuss this approximation in more detail in Appendix~\ref{appsec:params}.} and hence does not induce any spin-motion coupling. 
In a frame rotating at $\omega_s$, the total Hamiltonian including the ODF and the Raman beams is 
\begin{equation}
	\hat{H} = \sum_{j=1}^N B_j \hS_j^X + \sum_{n=1}^N \omega_n \hat{a}_n^\dag \hat{a}_n + 2\sum_{j=1}^N \delta_\mathrm{AC} \sin(\Delta k_x x_j + \Delta k_z \hat{z}_j -\mu_r t) \hS_j^Z.
\end{equation}
Here, $B_j$ is the effective two-photon Rabi frequency at the site of ion $j$. Assuming a beam waist $w\gg R$ for the Raman lasers, where $R$ is the crystal radius, we can approximate $B_j\approx B_0 -B_0r_j^2/w^2$. 

We now analyze the spin dynamics in a rotated spin space such that $\hSS_{\mcZ} \equiv -\hS_X$ and $\hSS_{\mcX} \equiv \hS_Z$. Further transforming to a frame rotating at $B_0$, the Hamiltonian in the rotated spin space is 
\begin{equation}
	\hat{H} =  \sum_{j=1}^N B_0\frac{r_j^2}{w^2}\hSS_j^{{\mcZ}} + \sum_{n=1}^N \omega_n \hat{a}_n^\dag \hat{a}_n 
	+ \sum_{j=1}^N \delta_\mathrm{AC} \sin(\Delta k_x x_j + \Delta k_z \hat{z}_j -\mu_r t) \left(\hSS_j^+ e^{-iB_0 t} + \hSS_j^- e^{iB_0 t}\right).
	\label{appeqn:rot_ham}
\end{equation}
We can now see the twofold role played by the Raman drive: By providing a splitting $2B_0$ between $\hSS_j^+$ and $\hSS_j^-$, it will enable retention of only desired interactions and enable ``rotating out'' unwanted interactions. Second, the beam waist $w$ acts as a knob to tune the single particle dispersion.

\subsection{Small angle approximation}

We now consider the sine function appearing in Eq.~(\ref{appeqn:rot_ham}). For three arguments $A=\Delta k_x x_j,B=\Delta k_z \hat{z}_j,C = -\mu_r t$, we can expand 
\begin{eqnarray}
	&&\sin(A+B+C) = \sin (A+B)\cos(C)  + \cos(A+B)\sin(C)
\end{eqnarray}
We assume that $A, B\ll 1$ and expand the relevant trigonometric functions in the small angle limit. The result, correct to second order in $A,B$, is 
\begin{eqnarray}
	\sin(A+B+C) \approx (A + B) \cos C + \left(1-\frac{A^2+B^2}{2}-AB\right)\sin C.  
	\label{appeqn:small_ang}
\end{eqnarray}

\subsection{Obtaining the two-channel model}

We express the resonance requirements as a sum of frequencies appearing in the argument of complex exponentials multiplying each interaction term. To do so, we first note that the motion along $x$ can be written as $x_j = r_j \cos(\omega_r t + \phi_j)=(r_j/2) (e^{i(\omega_r t+\phi_j)} + e^{-i(\omega_r t+\phi_j)})$, while the motion along $z$ can be expanded as
\begin{eqnarray}
\Delta k_z \hat{z}_j(t) = \sum_{n=1}^N \eta_n \mathcal{M}_{jn}\left(\hat{a}_n e^{-i\omega_n t}  + \hat{a}_n^\dagger e^{i \omega_n t} \right).
\label{appeqn:z_expansion_2}
\end{eqnarray}
This expansion assumes we have moved to an interaction picture with respect to the free phonon frequencies $\omega_n$. The homogeneous Raman drive sets a frequency $B_0$ for the spins as seen from Eq.~(\ref{appeqn:rot_ham}). Finally, the terms $\cos(-\mu_r t), \sin(-\mu_r t)$ can be expanded with complex exponentials of the form $e^{\pm i\mu_r t}$. For discussing the rotating wave approximations, we neglect the small contribution arising from the spatially inhomogeneous component of the Raman beams, i.e. we assume the beam waist $w\rightarrow\infty$. 

We tune $\mu_r$ to selectively induce a coupling between the spins ($B_0$), the drumhead c.m. mode ($\omega_1$) and the planar rotation ($\omega_r$). In particular, we adjust $\mu_r$ such that $B_0-\mu_r+\omega_1+\omega_r = \delta_1$. We will later consider the effect of off-resonant terms using realistic experimental parameters. Then, the only near-resonant term stems from the $AB\sin C$ type term in Eq.~(\ref{appeqn:small_ang}), and is of the form 

\begin{equation}
	\hat{H}_\text{I} = -\sum_{j=1}^N  \frac{G}{i\sqrt{N}} \td{r}_j \left(\hSS_j^- \hat{a}_1^\dag e^{i (\delta_1 t+\phi_j)} - \hSS_j^+ \hat{a}_1 e^{-i (\delta_1 t+\phi_j)}   \right),
	\label{eqn:jc}
\end{equation}
where $\td{r}_j = r_j/R$ is the ion radius normalized to the crystal radius $R$ (assuming a nearly circular crystal) and $G$ is given by 
\begin{equation}
	G = \frac{\delta_\mathrm{AC} \eta_1 \left(\Delta k_x R\right)}{4} .
\end{equation}
Here, $\eta_1 = \Delta k_z\sqrt{\hbar/(2m\omega_1)}$ is the Lamb-Dicke parameter for the c.m. mode. The quantity $\Delta k_x R$ can be thought of as an effective Lamb-Dicke parameter for the in-plane motion.

The above analysis was carried out assuming that the beam waist of the Raman lasers $w\rightarrow\infty$. Restoring a finite $w$ and performing a frame transformation for the c.m. mode, we arrive at the effective Hamiltonian 
\begin{eqnarray}
\hat{H}_\text{2ch} = \sum_{j=1}^N B_1 \td{r}_j^2 \hSS_j^{{\mcZ}} + \delta_1 \hat{a}_1^\dagger\hat{a}_1 
-\sum_{j=1}^N \frac{G}{i\sqrt{N}}\td{r}_j \left(\hSS_j^- \hat{a}_1^\dagger e^{i\phi_j}  - \hSS_j^+ \hat{a}_1 e^{-i\phi_j}
\right),
\label{eqn:ham_2ch_2}
\end{eqnarray}
where $B_1=B_0R^2/w^2$.

\subsection{Effective spin-exchange interaction}

We now eliminate the c.m. mode from Eq.~(\ref{eqn:ham_2ch_2}) using effective Hamiltonian theory~\cite{james2007CJP}. We express energy resonance requirements once again as complex exponentials by defining $\delta_1^j = \delta_1-B_1\td{r}_j^2$. Using a frame transformation for the spins and the c.m. mode, Eq.~(\ref{eqn:ham_2ch_2}) can be written as 
\begin{eqnarray}
\hat{H}_\text{I} = -\sum_{j=1}^N  \frac{G}{i\sqrt{N}} \td{r}_j \left(\hSS_j^- \hat{a}_1^\dag e^{i (\delta_1^j t+\phi_j)} - \hSS_j^+ \hat{a}_1 e^{-i (\delta_1^j t+\phi_j)}   \right).
\end{eqnarray}
By assuming that $\delta_1^j$ are large compared to the maximum interaction strength $G$, a spin-spin Hamiltonian can be derived using effective Hamiltonian theory. The result is 
\begin{equation}
	\hat{H}_\text{eff} = \sum_{j,k=1}^N \frac{G^2}{Nh(\delta_1^j,\delta_1^k)} \td{r}_j\td{r}_k [\hat{a}_1^\dag \hSS_j^- e^{i\left(\delta_1^j t+\phi_j\right)}, \hat{a}_1 \hSS_k^+ e^{-i\left(\delta_1^k t + \phi_k\right)}],
\end{equation}  
where $h(a,b)$ is the harmonic mean of $a,b$. The commutator evaluates to 
\begin{eqnarray}
	[\hat{a}_1^\dag \hSS_j^-,\hSS_k^{+}\hat{a}_{1}] = \left\{ \begin{array}{ll}
	-\hSS_j^-\hSS_k^+, \; j \neq k, \\
	-2\hat{a}_1^\dag \hat{a}_1 \hSS_j^{{\mcZ}} -\hSS_j^+\hSS_j^-, \; j = k.
	\end{array}
	\right.
\end{eqnarray}
The term containing $\hat{a}_1^\dag \hat{a}_1$ can be neglected if the c.m. mode is initially in the ground state such that $\ev{\hat{a}_1^\dag \hat{a}_1} = 0$. The effective Hamiltonian is therefore 
\begin{equation}
	\hat{H}_\text{eff} = - \sum_{j,k=1}^N \frac{G^2}{Nh(\delta_1^j,\delta_1^k)} \td{r}_j \td{r}_k \hSS_j^+ \hSS_k^- e^{-i\left[(\delta_1^j-\delta_1^k)t + (\phi_j-\phi_k)\right]}. 
\end{equation}
The time-dependence can be removed via a frame transformation to give 
\begin{eqnarray}
	\hat{H}_\text{eff} = \sum_{j=1}^N \left(B_1 \td{r}_j^2-\frac{G^2}{N\delta_1^j}\td{r}_j^2\right) \hSS_j^{{\mcZ}} -\sum_{j=1}^N\sum_{k\neq j} \frac{G^2}{N h(\delta_1^j,\delta_1^k)}\td{r}_j \td{r}_k \hSS_j^+ \hSS_k^- e^{-i(\phi_j-\phi_k)}.
\end{eqnarray}
Since $\delta_1 \gg G^2/\delta_1$ and $B_1$ is comparable to the latter frequency, as a first approximation we can assume $\delta_1^j\approx \delta_1^k = \delta_1$ and neglect site dependency in the denominators of the effective frequencies. This leads to the Hamiltonian for the one-channel model:
\begin{eqnarray}
	\hat{H}_\text{1ch} = K\sum_{j=1}^N \td{r}_j^2 \hSS_j^{{\mcZ}} -\frac{J}{N}\sum_{j=1}^N\sum_{k\neq j} \td{r}_j \td{r}_k \hSS_j^+ \hSS_k^- e^{-i(\phi_j-\phi_k)},
	\label{appeqn:ham_1ch}
\end{eqnarray}
where $J = G^2/\delta_1$ and $K = B_1-J/N$. We note that the total magnetization $\hat{\mathfrak{J}}^{{\mcZ}}=\sum_j \hSS_j^{{\mcZ}}$ is conserved by this Hamiltonian. 

\section{\label{appsec:params}Experimental parameters for implementation}

In this section, we provide experimental parameters for implementing our proposal. These parameters are based on settings used in the NIST Penning trap, where 2D crystals of tens to hundreds of ${}^9\mathrm{Be}^+$ ions are routinely prepared for quantum simulation and sensing. 

\subsection{\label{appsec:trap_params}Trapping parameters}

Two-dimensional crystals of ${}^9\mathrm{Be}^+$ ions are formed in the Penning trap by a combination of an electric quadrupole field providing axial confinement and a strong axial magnetic field $\mathcal{B}\approx 4.46 \;\mathrm{T}$ (cyclotron frequency $\omega_c/(2\pi)\approx 7.6 \;\mathrm{MHz}$) that aids in radial confinement. The spin-$1/2$ degree of freedom is encoded in two long-lived hyperfine levels of each ${}^9\mathrm{Be}^+$ ion, i.e. $\ket{\uparrow}\equiv \ket{2S_{1/2},1/2}$ and $\ket{\downarrow}\equiv \ket{2S_{1/2},-1/2}$. For a crystal with $N=200$ ions, the crystal radius is $R\sim 100\;\mu\mathrm{m}$. Here we consider two sets of trapping parameters. In case A, we will set the rotation frequency  to $\omega_r/(2\pi) = 180 \; \mathrm{kHz}$ and choose an axial trapping frequency, which is also the drumhead c.m. frequency, to be $\omega_1/(2\pi) = 1.59 \;\mathrm{MHz}$. In case B, we choose a faster rotating crystal with a higher axial trapping frequency, viz. $\omega_r/(2\pi) = 900 \; \mathrm{kHz}$ and  $\omega_1/(2\pi) = 3.42 \;\mathrm{MHz}$. In the following, we explicitly refer to the different cases when specifying parameters that are not the same in the two cases. While case A represents the commonly used trapping parameters, the reason we consider two sets of parameters will become clear in Appendix~\ref{appsec:unwanted}, where we show that certain off-resonant terms are only negligible for the faster rotating crystal, i.e. case B. 

In general, both the ODF beams and the Raman beams intersect the crystal plane at a non-zero angle. However, for the quantum simulation, we require that these beams have a radially varying intensity profile in the crystal plane. This can be achieved by using laser beams with elliptical beam waists, whose cross-section in the crystal plane will have a radial intensity profile.

\subsection{ODF interaction}

The optical dipole force is realized using a pair of lasers that intersect the crystal at approximately equal and opposite angles relative to the crystal plane, see Fig.~\ref{fig:fig1}(c). These lasers induce spatially varying AC Stark shifts on the pseudospin states by coupling these states to the $\ket{2P_{3/2}}$ manifold and give rise to a Hamiltonian of the form described in Eq.~(\ref{appeqn:ham_odf_tilted}).

From Fig.~\ref{fig:fig1}(c), the wavevectors of the ODF lasers are given by
\begin{eqnarray}
	\mbf{k}_u &=& (k\cos\theta) \hat{\mathbf{e}}_x - (k\sin\theta) \hat{\mathbf{e}}_z, \nonumber\\
	\mbf{k}_l &=& (k\cos(\theta+\delta\theta)) \hat{\mathbf{e}}_x + (k\sin(\theta+\delta\theta)) \hat{\mathbf{e}}_z,
\end{eqnarray}  
where $u,l$ denote the upper and lower ODF beams respectively and $k=\abs{\mbf{k}_u}=\abs{\mbf{k}_l}$. Denoting $k_x = k\cos\theta, k_z =k\sin\theta$, we can write 
\begin{eqnarray}
	\mbf{k}_u &=& k_x \hat{\mathbf{e}}_x - k_z \hat{\mathbf{e}}_z, \nonumber\\
	\mbf{k}_l &\approx& (k_x - k_z \delta\theta)\hat{\mathbf{e}}_x + (k_z + k_x\delta\theta) \hat{\mathbf{e}}_z.
\end{eqnarray}
The difference wavevector can then be expressed as 
\begin{equation}
	\Delta \mbf{k} = k_z\delta\theta \hat{\mathbf{e}}_x - (2k_z-k_x\delta\theta)\hat{\mathbf{e}}_z \equiv \Delta k_x \hat{\mathbf{e}}_x + \Delta k_z \hat{\mathbf{e}}_z.
\end{equation}

The wavevector magnitude  of each ODF laser is $k\approx 2 \times 10^7 \mathrm{m^{-1}}$. The Lamb-Dicke parameter along the axial direction is $\eta_1 = \Delta k_z l_1^{\mathrm{zp}}$, where $\Delta k_z \approx 2k\sin(\theta)$ is the difference wavevector along the $z$-direction and $l_1^{\mathrm{zp}}=\sqrt{\hbar/(2m\omega_1)}$ is the zero-point motion of the c.m. mode. Here, we have neglected the small correction to $\Delta k_z$ that arises from a non-zero misalignment $\delta\theta$. To obtain a value $\eta_1$, the ODF lasers must be oriented at angles $\pm\theta$ with respect to the crystal plane, with $\theta = \sin^{-1}(\eta_1/(2 k l_1^\mathrm{zp}))$. For $\eta_1\approx 0.3$, we find $\theta\approx 23.4^\circ$ for case A and $\theta\approx 35.7^\circ$ for case B.

An analogous small parameter along the $x$-direction is given by $\eta_x=\Delta k_x R$, where $\Delta k_x = k\delta\theta\sin\theta$ is the difference wavevector in the crystal plane. To obtain a value $\eta_x$, the required misalignment is given by $\delta\theta=\eta_x/(kR\sin\theta)$. For $\eta_x\approx 0.3$, we find $\delta\theta\approx 0.017^\circ$ for case A and $\delta\theta\approx 0.016^\circ$ for case B.  

We assume that $\delta_\mathrm{AC}/(2\pi)\approx 40\;\mathrm{kHz}$. Then, the interaction strength $G$ for the two-channel model is given by 
\begin{eqnarray}
\frac{G}{2\pi} = \frac{\delta_\mathrm{AC}\eta_1\eta_x}{4(2\pi)} = 900 \;\mathrm{Hz}.
\end{eqnarray}
By assuming a detuning of $\delta_1/(2\pi)= 2 \;\mathrm{kHz}$ from the c.m. mode, we can arrive at the effective one-channel model coupling strength $J$ as 
\begin{eqnarray}
\frac{J}{2\pi} = \frac{G^2}{\delta_1} = 405 \;\mathrm{Hz}.
\end{eqnarray}

In this parameter regime, the ratio $\delta_1^2/G^2\sim 5$, and therefore the simulation will have a small two-channel character to it in addition to the dominant one-channel model (see Fig.~\ref{fig:fig4}).

\subsection{Raman beams}

Two-photon Raman transitions between the pseudospin states can be engineered by introducing a pair of co-propagating Raman lasers that couple these states to the $\ket{2P_{3/2}}$ manifold in a far detuned regime. The purpose of the Raman beams is twofold. First, the spatially homogeneous two-photon Rabi frequency $B_0$ breaks the symmetry between the $\hSS^+$ and $\hSS^-$ terms, see Eq.~(\ref{appeqn:rot_ham}). We take this value to be $B_0/(2\pi)=10\;\mathrm{kHz}$. Second, the radially varying intensity of the Raman beams tunes the dispersion $K$ of the spins. We estimate the scale of the required beam waist $w$ as the value at which $K\approx J$, i.e. 
\begin{eqnarray}
K \approx \frac{B_0 R^2}{w^2} = J \implies w \approx 497 \;\mu\mathrm{m}.
\end{eqnarray}

An alternative mechanism to generate a Raman beam intensity gradient is by utilizing the Doppler shifts arising from the crystal rotation. Although the Raman beams are copropagating, their difference wavevector $\abs{\Delta k_R} \neq 0$  because of the frequency splitting $\omega_s$ of the spin states, i.e. $\abs{\Delta k_R}=\omega_s/c$, where $c$ is the speed of light in vacuum. Assuming the Raman beams are propagating in the $x-z$ plane and make an angle $\theta_R$ with the crystal plane, the Hamiltonian for a single ion interacting with the Raman beams is given by 
\begin{eqnarray}
\hat{H}_{R,j} = \omega_s \hS_j^Z +\frac{B_0}{2} \left( \hS_j^+ e^{i\left[\abs{\Delta \mbf{k}_R}\cos\theta_R x_j(t)-\Delta\omega_R t\right]} + \hS_j^- e^{-i\left[\abs{\Delta \mbf{k}_R}\cos\theta_R x_j(t)-\Delta\omega_R t\right]}\right).
\end{eqnarray}
In the Doppler free case, two-photon resonance is satisfied by setting $\Delta\omega_R = \omega_s$. In this case, transforming to a frame rotating at $\omega_s$ leads to the interaction Hamiltonian
\begin{eqnarray}
\hat{H}_{R,j}^I = \frac{B_0}{2} \left( \hS_j^+ e^{i\abs{\Delta \mbf{k}_R}\cos\theta_R x_j(t)} + \hS_j^- e^{-i\abs{\Delta \mbf{k}_R}\cos\theta_R x_j(t)}\right).
\end{eqnarray}
The Doppler shift is time modulated because $x_j(t)=r_j\cos(\omega_r t + \phi_j)$. Defining $\eta_j^R = \abs{\Delta \mbf{k}_R}\cos\theta_R r_j$, we can use the Jacobi-Anger expansion to write 
\begin{eqnarray}
e^{-i\eta_j^R \cos(\omega_r t + \phi_j) } = \sum_{n=-\infty}^{\infty} (-i)^n J_n(\eta_j^R) e^{-in\left(\omega_r t + \phi_j\right)},
\end{eqnarray}
where $J_n(x)$ is the $n^\mathrm{th}$ Bessel function of the first kind. The $n=0$ term is then given by 
\begin{eqnarray}
\hat{H}_{R,j}^{I,(0)} = B_0J_0(\eta_j^R)\hS_j^X \approx B_0 \left(1 - \frac{(\eta_j^R)^2}{4} \right) \hS_j^X = \left(B_0 - \frac{B_0\omega_s^2R^2}{4c^2}\td{r}_j^2 \right) \hS_j^X.
\end{eqnarray}
The approximation is valid for $(\eta_j^R)^2\ll 1$. To verify this, we consider the situation when the largest value of $\eta_j^R$ occurs, i.e., when $\theta_R=0$ and $r_j=R\sim100\;\mu$m, where $R$ is the crystal radius. Using $\omega_s/(2\pi) = 124\;$GHz, we estimate $\eta_j^R \approx 0.26$ and $(\eta_j^R)^2 \approx 0.067\ll 1$. The achievable value of $B_1$ in this case is given by 
\begin{eqnarray}
B_1 = \frac{\omega_s^2 R^2}{4 c^2}B_0 \approx 0.017B_0 \approx 2\pi \times 170 \; \mathrm{Hz},
\end{eqnarray}
where we have used $B_0/(2\pi)= 10\;\mathrm{kHz}$. Therefore, it appears that $B_1$ can be partially realized even without a beam waist by simply exploiting the crystal rotation.

We note that the dispersion $B_1$ arising from the Doppler shifts can also be achieved if Raman beams are replaced with a microwave drive that is tilted with respect to the spatial $z$-axis. Such an implementation may be simpler and will also eliminate off-resonant scattering from the Raman beams (see below). However, the long wavelength of microwaves precludes control of the beam waist at the $100 \;\mu$m level for additional tuning of $B_1$ that may be required for some aspects of our proposal.

\subsection{Decoherence from off-resonant light scattering}

We separately estimate the contributions from the ODF beams and the Raman beams and find them to be~\cite{uys2010PRL,carterNotes} 
\begin{eqnarray}
\frac{\Gamma_\text{ODF}}{2\pi} \approx 38 \;\text{Hz},\; \frac{\Gamma_\text{Raman}}{2\pi} \approx 15 \;\text{Hz}.  
\end{eqnarray}
The total decoherence rate is then $\Gamma_\text{tot}/(2\pi) = (\Gamma_\text{ODF} + \Gamma_\text{Raman})/(2\pi) \approx 53 \; \text{Hz}$. 

With the chosen parameters, we estimate the typical time up to which the simulation can be run as $Jt \approx J(1/\Gamma_\text{tot}) \sim 7.6$. 
As we mention in the Main Text, the ratio of the coherent interaction to the decoherence rate can be enhanced by a number of means, including choosing a different ion species and transition, enhancing coherent coupling by parametric amplification and by working at a different ODF operating point. 

We note that because of the multilevel structure of the electronic excited states, the full decoherence model for the ODF beams and for the Raman beams contains a number of independent decay channels with nontrivial rates and jump operators that are strongly modified by multilevel interference effects. Although we consider the full decoherence model~\cite{uys2010PRL,carterNotes}, we have only roughly estimated the decoherence rate by inspecting analytical equations for the rate of decay of individual spin components. In the future, a detailed study of the impact of decoherence can be performed for specific experimental settings by including the full decoherence model in the numerical simulation.

\section{\label{appsec:rwa}Rotating-wave approximations}

In deriving the two-channel model, we assumed that all other terms in the expansion Eq.~(\ref{appeqn:small_ang}) could be considered rapidly oscillating. We now check whether this is true by first listing the various resonance conditions and estimating their values using real experimental parameters.

We first consider the term of the form $AB\sin C\hSS_j^-$, for which the possible arguments for the complex exponentials are (up to an overall sign, which gives the hermitian conjugate process with $\hSS_j^+$)
\begin{eqnarray}
	\allowdisplaybreaks
    T_{11} \equiv 	B_0 - \mu_r + \omega_n + \omega_r &=& \delta_n \nonumber\\
	T_{12} \equiv B_0 - \mu_r + \omega_n - \omega_r &=& \delta_n -2\omega_r \nonumber\\
	T_{13} \equiv B_0 - \mu_r - \omega_n + \omega_r &=& \delta_n -2\omega_n \nonumber\\
	T_{14} \equiv B_0 - \mu_r - \omega_n - \omega_r &=& \delta_n -2\omega_n -2\omega_r \nonumber\\	T_{15} = B_0 + \mu_r + \omega_n + \omega_r &=& \delta_n +2\mu_r \nonumber\\	
	T_{16} \equiv B_0 + \mu_r + \omega_n - \omega_r &=& \delta_n +2\mu_r -2\omega_r \nonumber\\	
	T_{17} \equiv B_0 + \mu_r - \omega_n + \omega_r &=& \delta_n +2\mu_r -2\omega_n \nonumber\\	
	T_{18} \equiv B_0 + \mu_r - \omega_n - \omega_r &=& -\delta_n + 2B_0.
	\label{eqn:absinC_res}
\end{eqnarray}

We next consider the term of the form $B^2 \sin C\hSS_j^-$. Here, we have 
\begin{eqnarray}
	\allowdisplaybreaks	
	T_{21} \equiv B_0 - \mu_r + \omega_n + \omega_k &=& \delta_1 +(\omega_n+\omega_k) - (\omega_1+\omega_r) \nonumber\\
	T_{22} \equiv B_0 - \mu_r + \omega_n - \omega_k &=& \delta_1 +(\omega_n-\omega_k) - (\omega_1+\omega_r) \nonumber\\
	T_{23} \equiv B_0 - \mu_r - \omega_n + \omega_k &=& \delta_1 +(\omega_k-\omega_n) - (\omega_1+\omega_r) \nonumber\\
	T_{24} \equiv B_0 - \mu_r - \omega_n - \omega_k &=& \delta_1 -(\omega_n+\omega_k) - (\omega_1+\omega_r) \nonumber\\	
	T_{25} \equiv B_0 + \mu_r + \omega_n + \omega_k &=& \delta_1 +2\mu_r  +(\omega_n+\omega_k) - (\omega_1+\omega_r)\nonumber\\	
	T_{26} \equiv B_0 + \mu_r + \omega_n - \omega_k &=& \delta_1 +2\mu_r  +(\omega_n-\omega_k) - (\omega_1+\omega_r) \nonumber\\	
	T_{27} \equiv B_0 + \mu_r - \omega_n + \omega_k &=& \delta_1 +2\mu_r  +(\omega_k-\omega_n) - (\omega_1+\omega_r) \nonumber\\	
	T_{28} \equiv B_0 + \mu_r - \omega_n - \omega_k &=& \delta_1 +2\mu_r  -(\omega_n+\omega_k) - (\omega_1+\omega_r).
\end{eqnarray}

Next, we consider terms of the form $A^2 \sin C \hSS_j^-$. For these terms, we get 
\begin{eqnarray}
	\allowdisplaybreaks
	T_{31} \equiv B_0 - \mu_r + \omega_r + \omega_r &=& \delta_n +\omega_r - \omega_n \nonumber\\
	T_{32} \equiv B_0 - \mu_r + \omega_r - \omega_r &=& \delta_n -\omega_r - \omega_n \nonumber\\
	T_{33} \equiv B_0 - \mu_r - \omega_r + \omega_r &=& \delta_n -\omega_r - \omega_n \nonumber\\
	T_{34} \equiv B_0 - \mu_r - \omega_r - \omega_r &=& \delta_n -3\omega_r -\omega_n \nonumber\\	
	T_{35} \equiv B_0 + \mu_r + \omega_r + \omega_r &=& \delta_n +2\mu_r +\omega_r - \omega_n \nonumber\\	
	T_{36} \equiv B_0 + \mu_r + \omega_r - \omega_r &=& \delta_n +2\mu_r -\omega_r - \omega_n \nonumber\\	
	T_{37} \equiv B_0 + \mu_r - \omega_r + \omega_r &=& \delta_n +2\mu_r -\omega_r - \omega_n \nonumber\\	
	T_{38} \equiv B_0 + \mu_r - \omega_r - \omega_r &=& \delta_n +2\mu_r -3\omega_r -\omega_n.
\end{eqnarray}

Now, we consider terms of the form $\sin C \hSS_j^-$. Here, the complex exponentials are simply $T_{41}\equiv B_0+\mu_r$ and $T_{42}\equiv B_0-\mu_r$, which respectively evaluate to $\delta_n+2\mu_r-\omega_n-\omega_r$ and $\delta_n-\omega_n-\omega_r$. 
 	
We now turn to the term of the form $B\cos C \hSS_j^-$. Here, we have the following detunings 
\begin{eqnarray}
	\allowdisplaybreaks	
	T_{51} \equiv B_0 - \mu_r + \omega_n &=& \delta_n - \omega_r \nonumber \\
	T_{52} \equiv B_0 - \mu_r - \omega_n &=& \delta_n - 2\omega_n - \omega_r \nonumber \\
	T_{53} \equiv B_0 + \mu_r + \omega_n &=& \delta_n +2\mu_r -\omega_r \nonumber \\
	T_{54} \equiv B_0 + \mu_r - \omega_n &=& \delta_n +2\mu_r -2\omega_n -\omega_r.
\end{eqnarray}

Finally, we consider the term of the form $A\cos C \hSS_j^-$. Here, we get 
\begin{eqnarray}
	\allowdisplaybreaks	
	T_{61} \equiv B_0 - \mu_r + \omega_r &=& \delta_n - \omega_n \nonumber \\
	T_{62} \equiv B_0 - \mu_r - \omega_r &=& \delta_n - 2\omega_r - \omega_n \nonumber \\
	T_{63} \equiv B_0 + \mu_r + \omega_r &=& \delta_n +2\mu_r -\omega_n \nonumber \\
	T_{64} \equiv B_0 + \mu_r - \omega_r &=& \delta_n +2\mu_r -2\omega_r -\omega_n.
\end{eqnarray}

In order to estimate these expressions, we compute the equilibrium crystal structure for cases A and B, and obtain the drumhead mode spectrum to obtain the frequencies $\omega_n$. In Table~\ref{tab:rwa_values}, we provide the maximum and minimum values possible for each of these terms and for cases A and B. If the maximum and minimum values are large compared to $\abs{\delta_1}$ and have the same sign, then there are no accidental resonances and the terms can be safely neglected.  We have excluded the c.m. term when evaluating the range of $T_{11}$ as this is precisely the term of interest with $\delta_1/(2\pi)\leq 2\;\mathrm{kHz}$. Although at first glance all of these terms appear to be far off-resonant, we explore the impact of low-order off-resonant terms in a quantitative manner in the next section. 


 \renewcommand{\arraystretch}{1.5}
\begin{table}[!tb]
    \centering
    \begin{tabular}{|c|c|c|}
\hline
term    &   case A  & case B \\
\hline
$T_{11}$    &   (-524, -19)	&	(-505, -21)   \\
$T_{12}$    &   (-884, -358)	&	(-2305, -1798) \\
$T_{13}$    &   (-3178, -2652)	&	(-6849, -6343) \\
$T_{14}$    &   (-3538, -3012)	&	(-8649, -8143) \\
$T_{15}$    &   (3032, 3558)	&	(8163, 8669) \\
$T_{16}$    &   (2672, 3198)	&	(6363, 6869) \\
$T_{17}$    &   (378, 904)	&	(1818, 2325) \\
$T_{18}$    &   (18, 544)	&	(18, 525) \\
\hline
$T_{21}$    &   (360, 1412)	&	(1514, 2528) \\
$T_{22}$    &   (-2294, -1242)	&	(-4830, -3817) \\
$T_{23}$    &   (-2294, -1242)	&	(-4830, -3817) \\
$T_{24}$    &   (-4948, -3896)	&	(-11175, -10162) \\
$T_{25}$    &   (3916, 4968)	&	(10182, 11195) \\
$T_{26}$    &   (1262, 2314)	&	(3837, 4850) \\
$T_{27}$    &   (1262, 2314)	&	(3837, 4850) \\
$T_{28}$    &   (-1392, -340)	&	(-2508, -1494) \\
\hline
$T_{31}$    &   (-1408, -1408)	&	(-2524, -2524) \\
$T_{32}$    &   (-1768, -1768)	&	(-4324, -4324) \\
$T_{33}$    &   (-1768, -1768)	&	(-4324, -4324) \\
$T_{34}$    &   (-2128, -2128)	&	(-6124, -6124) \\
$T_{35}$    &   (2148, 2148)	&	(6144, 6144) \\
$T_{36}$    &   (1788, 1788)	&	(4344, 4344) \\
$T_{37}$    &   (1788, 1788)	&	(4344, 4344) \\
$T_{38}$    &   (1428, 1428)	&	(2544, 2544) \\
\hline
$T_{41}$    &   (1788, 1788)	&	(4344, 4344) \\
$T_{42}$    &   (-1768, -1768)	&	(-4324, -4324) \\
\hline
$T_{51}$    &   (-704, -178)	&	(-1405, -898) \\
$T_{52}$    &   (-3358, -2832)	&	(-7749, -7243) \\
$T_{53}$    &   (2852, 3378)	&	(7263, 7769) \\
$T_{54}$    &   (198, 724)	&	(918, 1425) \\
\hline
$T_{61}$    &   (-1588, -1588)	&	(-3424, -3424) \\
$T_{62}$    &   (-1948, -1948)	&	(-5224, -5224) \\
$T_{63}$    &   (1968, 1968)	&	(5244, 5244) \\
$T_{64}$    &   (1608, 1608)	&	(3444, 3444) \\
\hline
\end{tabular}
\caption{Minimum and maximum values of various resonance conditions for both trapping parameters, viz. case A and case B. The frequencies are reported in units of kilohertz.}
\label{tab:rwa_values}
\end{table}

\section{\label{appsec:unwanted}One-channel model and off-resonant terms}

For simulating the one-channel model, we assumed that $\delta_{1}/(2\pi)=2\;\mathrm{kHz}$ constitutes a large detuning based on which the c.m. mode can be adiabatically eliminated. The resulting interaction is fourth order in the small parameters $A,B$ of the small angle expansion Eq.~(\ref{appeqn:small_ang}). This raises the question whether some of the other terms in Eq.~(\ref{appeqn:small_ang}), although off-resonant, could potentially compete with the effective one-channel interactions because they are of lower order in the small parameters. In this section, we consider the effect of terms that are zeroth and first order in the small parameters on the effective dynamics of the one-channel model. We also estimate off-resonant effects that arise from the second-order term $AB\sin C$ since some terms in this group are not very far off-resonant.

\subsection{$\sin C$ term}

This term is of the form 
\begin{eqnarray}
\hat{H}_4 = \sum_j \frac{\delta_\text{AC}}{2i}\left(\hSS_j^+ e^{-i (\mu_r+B_j) t} + \hSS_j^- e^{-i (\mu_r-B_j) t} \right) + \text{h.c.}
\end{eqnarray}

Using effective Hamiltonian theory (EHT) and assuming $B_j\ll \mu_r$ and $B_j\approx B_0$ leads to 
\begin{eqnarray}
\hat{H}_{4,\text{eff}} = \sum_j \frac{\delta_\text{AC}^2 B_0}{\mu_r^2} \hSS_j^z.
\end{eqnarray}
This term represents a small AC Stark shift that leads to collective spin precession at a frequency $\sim 5 \;\mathrm{Hz}$ for case A and $\sim 0.9\;\mathrm{Hz}$ for case B. Hence, its effect can be considered negligible.

\subsection{$A\cos C$ term}

This term is of the form 
\begin{eqnarray}
\hat{H}_6 = \sum_j \frac{\delta_\text{AC}\eta_x \td{r}_j}{4}
&&\left(\hSS_j^+ e^{-i [(\mu_r+\omega_r+B_j) t + \phi_j]} + \hSS_j^+ e^{-i [(\mu_r-\omega_r+B_j) t - \phi_j]} \right.\nonumber\\
&&\left. + \hSS_j^- e^{-i [(\mu_r+\omega_r-B_j) t + \phi_j]} + \hSS_j^- e^{-i [(\mu_r-\omega_r-B_j) t - \phi_j]} \right) + \text{h.c.}
\end{eqnarray}
Using EHT leads to 
\begin{eqnarray}
\hat{H}_{6,\text{eff}} = -\sum_j \frac{\delta_\text{AC}^2\eta_x^2 \td{r}_j^2}{8} \left[ \frac{1}{\mu_r+\omega_r+B_j} + \frac{1}{\mu_r-\omega_r+B_j} - \frac{1}{\mu_r+\omega_r-B_j} -\frac{1}{\mu_r-\omega_r-B_j} \right] \hSS_j^z.
\end{eqnarray}
Assuming $B_j\sim B_0 \ll \omega_r,\mu_r$, we get 
\begin{eqnarray}
\hat{H}_{6,\text{eff}} = \sum_j \frac{\delta_\text{AC}^2 B_0 \eta_x^2 \td{r}_j^2}{4} \left[\frac{1}{(\mu_r+\omega_r)^2} + \frac{1}{(\mu_r-\omega_r)^2}\right] \hSS_j^z.
\end{eqnarray}
This term results in a radius-dependent AC Stark shift. We estimate its maximal value by setting $\td{r}_j=1$, for which we find a precession frequency $\sim 0.5\;\mathrm{Hz}$ for case A and $\sim 0.04\;\mathrm{Hz}$ for case B. These are very small compared to the dispersion generated by the Raman beam waist and hence we neglect these terms.

\subsection{$B\cos C$ term}

This term is of the form 
\begin{eqnarray}
\hat{H}_5 = \sum_{j,n} \frac{\delta_\text{AC}\eta_n\mathcal{M}_{jn}}{2}
&&\left( \hSS_j^+ \hat{a}_n e^{-i(\mu_r+B_0+\omega_n)t} + \hSS_j^+ \hat{a}_n^\dag e^{-i(\mu_r+B_0-\omega_n)t}  \right. \nonumber\\
&&\left. +  \hSS_j^- \hat{a}_n e^{-i(\mu_r-B_0+\omega_n)t} + \hSS_j^- \hat{a}_n^\dag e^{-i(\mu_r-B_0-\omega_n)t}  \right) + \;\text{h.c.}
\end{eqnarray}
For applying EHT, we evaluate the following commutators
\begin{eqnarray}
\left[\hSS_j^+\hat{a}_n, \hSS_k^-\hat{a}_m^\dag \right] &=& 2\hat{a}_m^\dag \hat{a}_n \hSS_j^z \delta_{jk} + \delta_{nm}\hSS_j^+\hSS_k^-, \nonumber\\
\left[\hSS_j^+\hat{a}_n^\dag, \hSS_k^-\hat{a}_m \right] &=& 2\hat{a}_m \hat{a}_n^\dag \hSS_j^z \delta_{jk} - \delta_{nm}\hSS_j^+\hSS_k^-, \nonumber\\
\left[\hSS_j^-\hat{a}_n, \hSS_k^+\hat{a}_m^\dag \right] &=& -2\hat{a}_m^\dag \hat{a}_n \hSS_j^z \delta_{jk} + \delta_{nm}\hSS_j^-\hSS_k^+, \nonumber\\
\left[\hSS_j^-\hat{a}_n^\dag, \hSS_k^+\hat{a}_m \right] &=& -2\hat{a}_m \hat{a}_n^\dag \hSS_j^z \delta_{jk} - \delta_{nm}\hSS_j^-\hSS_k^+.
\end{eqnarray}
We only consider the effective role of each term in the parenthesis independently. 
With the assumption of ground state cooling, we make the replacement $\hat{a}_m^\dag \hat{a}_n\rightarrow 0$ and $\hat{a}_m \hat{a}_n^\dag \rightarrow \delta_{nm}$. We then have the effective Hamiltonian 
\begin{eqnarray}
\hat{H}_{5,\text{eff}} = &&-\sum_j \sum_n \frac{\delta_\text{AC}^2\eta_n^2\mathcal{M}_{jn}^2}{4}\left[\frac{1}{\mu_r+B_0+\omega_n} + \frac{1}{\mu_r+B_0-\omega_n}
-\frac{1}{\mu_r-B_0+\omega_n}
-\frac{1}{\mu_r-B_0-\omega_n}\right]\hSS_j^z \nonumber\\
&&-\sum_{j \neq k} \sum_n \frac{\delta_\text{AC}^2\eta_n^2\mathcal{M}_{jn}\mathcal{M}_{kn}}{4}\left[\frac{1}{\mu_r+B_0+\omega_n} - \frac{1}{\mu_r+B_0-\omega_n}
+\frac{1}{\mu_r-B_0+\omega_n}
-\frac{1}{\mu_r-B_0-\omega_n}\right] \hSS_j^+ \hSS_k^-.
\end{eqnarray}
As a first approximation, we neglect terms with $\mu_r+\omega_n$ in the denominator since they are small compared with terms that have $\mu_r-\omega_n$ in the denominator. Using the fact that $B_0\ll \mu_r-\omega_n$, we arrive at the approximate effective Hamiltonian
\begin{eqnarray}
\hat{H}_{5,\text{eff}} = \sum_{j,k} J_{5,jk}\hSS_j^+\hSS_k^-,
\end{eqnarray}
where the interaction matrix $J_5$ has elements given by 
\begin{eqnarray}
J_{5,jk} = \left\{ \begin{array}{cc}
\sum_n \dfrac{\delta_\text{AC}^2 B_0 \eta_n^2\mathcal{M}_{jn}^2}{2(\mu_r-\omega_n)^2},     &  j=k\\
 \sum_n \dfrac{\delta_\text{AC}^2\eta_n^2\mathcal{M}_{jn}\mathcal{M}_{kn}}{2(\mu_r-\omega_n)},    &    j\neq k. 
\end{array}
\right.
\end{eqnarray}

The effective Hamiltonian $\hat{H}_{5,\text{eff}}$ mediates achiral spin-exchange type interactions that directly compete with the chiral spin-exchange that we wish to engineer. We study the impact of this term numerically in Fig.~\ref{fig:unwanted} for case A and case B using mean-field theory. We find that in case A, this interaction causes $\abs{\psi(t)}$ to rapidly decay toward zero on short timescales whereas in case B, the impact of this term is rather small. From the form of the elements in the coupling matrix $J_5$, a larger rotation frequency $\omega_r$ increases the denominator, i.e. makes these terms smaller and hence their impact is smaller in case B.

\subsection{$AB\sin C$ term}

The interaction giving rise to the two-channel model is present in this second-order term and was discussed previously. Here, we estimate the contribution of other terms present in this interaction since some of them are not very far off-resonant. 

This term is of the form 
\begin{eqnarray}
\hat{H}_1 = -\sum_{j,n} \frac{\delta_\text{AC}\eta_x \td{r}_j \eta_n \mathcal{M}_{jn}}{4i}
&&\left(\hSS_j^+\hat{a}_n^\dag e^{-i\left[(\mu_r-\omega_n-\omega_r+B_0)t-\phi_j\right]}
+ \hSS_j^+\hat{a}_n^\dag e^{-i\left[(\mu_r-\omega_n+\omega_r+B_0)t+\phi_j\right]} \right.\nonumber\\
&&\left. 
+ \hSS_j^-\hat{a}_n^\dag e^{-i\left[(\mu_r-\omega_n-\omega_r-B_0)t-\phi_j\right]}
+ \hSS_j^-\hat{a}_n^\dag e^{-i\left[(\mu_r-\omega_n+\omega_r-B_0)t+\phi_j\right]}
\right) + \; \text{h.c.}
\end{eqnarray}
In writing the above equation, we have already ignored the terms that have exponentials containing the combination $\mu_r+\omega_n$. Following a calculation similar to the $B\cos C$ term, we find an effective Hamiltonian given by 
\begin{eqnarray}
\hat{H}_{1,\text{eff}} = \sum_{j,k} J_{11,jk}\hSS_j^+\hSS_k^- 
+ \sum_{j,k} J_{12,jk}\hSS_j^+\hSS_k^-,
\end{eqnarray}
where the interaction matrices $J_{11},J_{12}$ have elements given by 
\begin{eqnarray}
&&J_{11,jk} = \left\{ \begin{array}{cc}
\sum_n \dfrac{\delta_\text{AC}^2\eta_x^2 \td{r}_j^2\eta_n^2\mathcal{M}_{jn}^2}{16(\delta_n-2B_0)},     & j=k \\
-\sum_n \dfrac{\delta_\text{AC}^2\eta_x^2 \td{r}_j \td{r}_k\eta_n^2\mathcal{M}_{jn} \mathcal{M}_{kn}}{16}\left[\dfrac{1}{\delta_n-2B_0} + \dfrac{1}{\delta_n-2\omega_r} \right]e^{i(\phi_j-\phi_k)},     & j\neq k,
\end{array}
\right. \nonumber\\
&&J_{12,jk} = \left\{ \begin{array}{cc}
-\sum_n \dfrac{\delta_\text{AC}^2\eta_x^2 \td{r}_j^2\eta_n^2\mathcal{M}_{jn}^2}{16\delta_n},     & j=k \\
-\sum_n \dfrac{\delta_\text{AC}^2\eta_x^2 \td{r}_j \td{r}_k\eta_n^2\mathcal{M}_{jn} \mathcal{M}_{kn}}{16}\left[\dfrac{1}{\delta_n-2\omega_r} + \dfrac{1}{\delta_n} \right]e^{-i(\phi_j-\phi_k)},     & j\neq k.
\end{array}
\right.
\end{eqnarray}

The $J_{12}$ matrix now represents the chiral spin-exchange interactions arising from all drumhead modes. On the other hand, the $J_{11}$ matrix describes anti-chiral interactions mediated by these modes. We estimate the impact of these terms numerically in Fig.~\ref{fig:unwanted} and find that they do not significantly impact the one-channel model dynamics in both cases A and B.

\begin{figure}[!ht]
    \centering
    \includegraphics[width=0.75\textwidth]{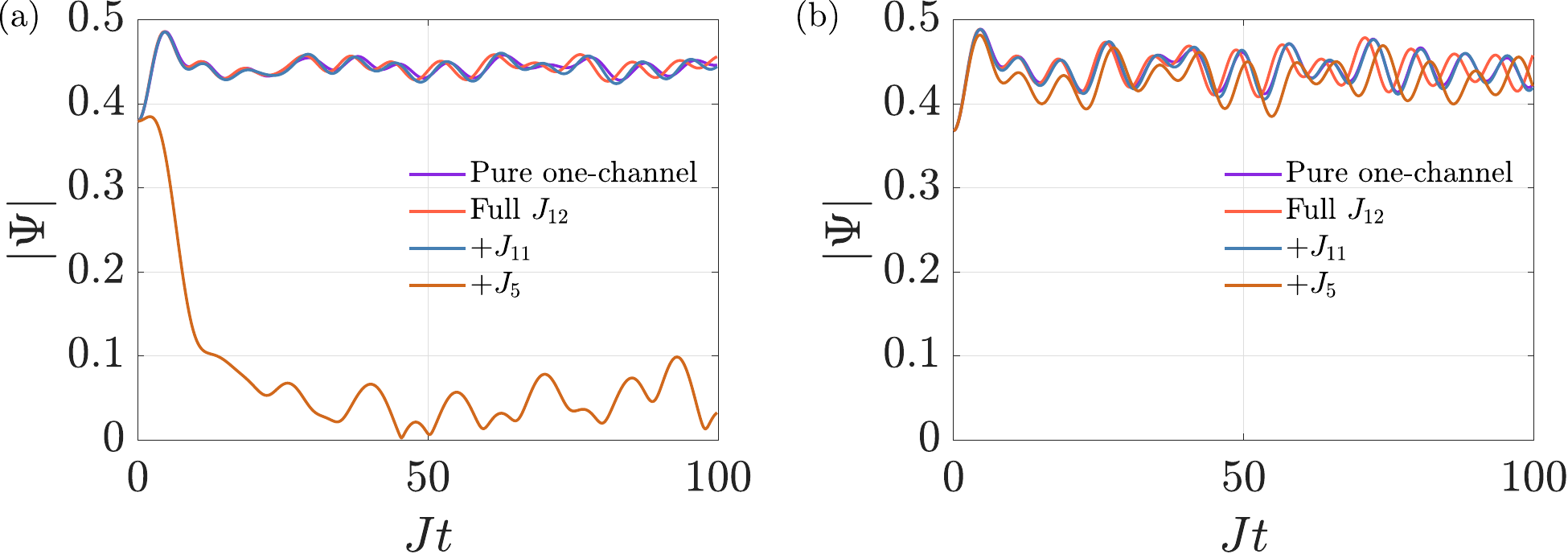}
    \caption{Impact of off-resonant interactions on the one-channel model dynamics for crystals formed under two different trapping parameters, (a) case A and (b) case B (see Appendix~\ref{appsec:trap_params}). In both cases, we progressively add terms to the pure one-channel model and study their impact. We add the following terms in the specified order: Chiral spin-exchange by all modes ($J_{12}$), anti-chiral spin-exchange by all modes ($J_{11}$) and achiral spin-exchange by all modes ($J_5$). Here, we start from a BCS-like initial state [top panel of Fig.~\ref{fig:fig2}(a)] and use experimental parameters discussed in Appendix~\ref{appsec:params}. We have used $B_1\approx J$, where the value of $J/(2\pi)\approx 405 \;\mathrm{Hz}$.}
    \label{fig:unwanted}
\end{figure}

Our study of off-resonant interactions shows that trapping parameters can be found where the impact of these terms can be neglected for realizing the one-channel model. In the case of the two-channel model, preliminary numerical results indicate that, even in case A, the adverse impact of off-resonant terms on short timescales ($Gt\lesssim 30$) is reduced as the c.m. mode is brought near resonance, suggesting that two-channel dynamics could be probed even with case A trapping parameters. The pure one-channel and pure two-channel model results presented in the Main Text have been computed using the case A equilibrium crystal. We have checked that all the results for the case B equilibrium crystal are qualitatively similar to the case A results when off-resonant terms are neglected. 


We end this section by noting that we have not extensively considered the impact of cross-talk between different terms as well as the role of terms at third and higher order in the small parameters, because of the sheer number of terms and their combinations. Their impact and the parameter regimes where they are negligible could potentially be explored directly on the quantum simulator.



\section{\label{appsec:num}Numerical methods}

While the classification of dynamical phases is based on a mean-field study in the thermodynamic limit, the finite size of our system has motivated us to consider beyond-mean-field techniques for numerical solution. Here, we present the mean-field equations for the one-channel and two-channel models, discuss how we include the quantum noise of the initial state using the discrete truncated Wigner approximation (dTWA) method, and benchmark the performance of dTWA using special crystals where the geometry allows for exact numerical solution of the Schrodinger equation. 

\subsection{Mean-field equations of motion for one-channel and two-channel models}

In the mean-field theory for the one-channel model, we replace the spin vector operator $\boldsymbol{\hSS}_j$ at each site $j$ by a vector of $c$-numbers $\boldsymbol{\cSS}_j \equiv \left(\cSS_j^{{\mcX}},\cSS_j^{{\mcY}},\cSS_j^{{\mcZ}}\right)$. The commutation relations are replaced by Poisson brackets, $\left\{\cSS_j^a,\cSS_k^b\right\}=i\epsilon_{abc}\cSS_j^c \delta_{jk}$. Defining $\cSS_j^\pm = \cSS_j^{{\mcX}} \pm i \cSS_j^{{\mcY}}$, the dynamics under Hamiltonian~(\ref{appeqn:ham_1ch}) are given by 
\begin{eqnarray}
    \frac{d}{dt}\cSS_j^+ &=& iK \td{r}_j^2 \cSS_j^+
    +i J\td{r}_j \cSS_j^{{\mcZ}} e^{i\phi_j}  \Psi_j^*, \nonumber\\
    \frac{d}{dt} \cSS_j^{{\mcZ}} &=& 
    - J\td{r}_j \mathrm{Im}\left[\cSS_j^+ e^{-i\phi_j} \Psi_j\right],   
\end{eqnarray}
where $\Psi_j = (2/N) \sum_{k \neq j} \td{r}_k \cSS_k^- e^{i\phi_k}$.

To obtain the mean-field equations for the two-channel model governed by Hamiltonian~(\ref{eqn:ham_2ch_2}), we additionally replace the annihilation operator $\hat{a}_1$ of the c.m. mode by a complex amplitude $\alpha_1$. The resulting equations are 
\begin{eqnarray}
    \frac{d}{dt} \cSS_j^+ &=& i B_1 \td{r}_j^2 \cSS_j^+
    + \frac{2G}{\sqrt{N}} \alpha_1^* \cSS_j^{{\mcZ}} \td{r}_j  e^{i\phi_j}, \nonumber\\
    \frac{d}{dt} \cSS_j^{{\mcZ}} &=& -\frac{2G}{\sqrt{N}}\td{r}_j\mathrm{Re}\left[s_j^+ \alpha_1 e^{-i\phi_j} \right], \nonumber\\
    \frac{d}{dt}\alpha_1 &=& -i\delta_1 \alpha_1
    +\frac{\sqrt{N} G}{2} \Psi,
\end{eqnarray}
with $\Psi = (2/N) \sum_j \td{r}_j \cSS_j^- e^{i\phi_j}$.

For mean-field simulations, the above equations are numerically evolved starting from initial conditions where the $c$-numbers are assigned the expectation values of the corresponding quantum operators in the initial state.

\subsection{Accounting for initial quantum noise}

For finite size systems, quantum corrections to the mean-field dynamics become important. To explore their effects, we simulate the effects of the quantum noise of the initial state by evolving several trajectories under the mean-field equations starting from randomly drawn initial conditions. For the spin degrees of freedom, we first find the mean spin direction $\hat{\mathbf{e}}_j^\parallel$ in the initial state. Next, we identify two mutually orthogonal spin directions, $\hat{\mathbf{e}}_j^{\perp,1}, \hat{\mathbf{e}}_j^{\perp,2}$, in the plane perpendicular to the mean spin. The initial spin vector can then be written as 
\begin{eqnarray}
    \boldsymbol{\cSS}_j = \cSS_j^\parallel \hat{\mathbf{e}}_j^\parallel + \cSS_j^{\perp,1}  \hat{\mathbf{e}}_j^{\perp,1} + \cSS_j^{\perp,2} \hat{\mathbf{e}}_j^{\perp,2}.
\end{eqnarray}
For mean-field simulations, we set $\cSS_j^\parallel=1/2, \cSS_j^{\perp,1} = \cSS_j^{\perp,2} = 0$. To go beyond mean-field, we use the prescription of the discrete truncated Wigner approximation~\cite{scachenmayer2015PRX}, according to which $\cSS_j^{\perp,1}$ and $\cSS_j^{\perp,2}$ are independently and randomly chosen to be $\pm 1/2$ with equal probability.

In the case of the two-channel model, we additionally draw the complex amplitude $\alpha_1$ from the Wigner distribution of the initial state of the c.m. mode, which we always assume is the motional ground state in this work. Therefore, the real and imaginary parts of $\alpha_1$ are independently drawn from a Gaussian distribution with zero mean and a variance of $1/4$.

\subsection{Benchmarking the dTWA results}

In the case of a real crystal in a Penning trap, the triangular lattice is only approximate and hence every ion is typically at a slightly different radius from the trap center. This makes an exact solution of the Schrodinger equation subject to Hamiltonian~(\ref{appeqn:ham_1ch}) exponentially hard. Therefore, in order to test the reliability of the dTWA results, we test this technique on a hypothetical ideal crystal made of $M$ concentric rings of ions, for which an exact numerical solution to the Schrodinger equation is feasible. We assume that the number of ions $N_m$ in ring $m=1,\ldots, M$ is given by $N_m = 6(m-1) + \delta_{m,1}$, which roughly mimics the lattice structure of closed-shell Penning trap crystals. Here, the first ring $m=1$ is taken to be the single ion at the crystal center. For simplicity, we assume that the radius of the rings $r_m$ grows linearly with ring index $m$ with $r_1=0$. 

For this model, we can define total angular momentum operators for each ring, $\hat{\mfJ}_m^\pm, \hat{\mfJ}_m^{\mcZ}$ as 
\begin{eqnarray}
    \hat{\mfJ}_m^\pm = \sum_{j \;\in \mathrm{\; ring\; } m} \hSS_j^\pm e^{\mp i\phi_j}, \; \hat{\mfJ}_m^{{\mcZ}} = \sum_{j \;\in \mathrm{\; ring\; } m} \hSS_j^{{\mcZ}}.
\end{eqnarray}
These operators are readily seen to obey the usual angular momentum commutation relations. In terms of these operators, the one-channel model~(\ref{appeqn:ham_1ch}) can be expressed as 
\begin{eqnarray}
    \hat{H}_\mathrm{eff} = B_1\sum_{m=1}^M\td{r}_m^2\hat{\mfJ}_m^{{\mcZ}} - 
    \frac{J}{N} \sum_{m,m'=1}^M \td{r}_m \td{r}_{m'} \hat{\mfJ}_m^+ \hat{\mfJ}_{m'}^-.
\end{eqnarray}
For a crystal with $M$ rings, the total number of ions is $N(M) = 1+3M(M-1)$. The computational complexity is significantly reduced in the total angular momentum picture, because, for the initial states we consider, we only need to track the fully symmetric subspace of each ring. Therefore, the number of basis states in each ring is reduced from $2^{N_m}$ to $N_m+1$, thereby enabling the rapid simulation of exact dynamics for crystals with up to $M=5$ rings ($N(5)=61$ ions). Figure~\ref{appfig:exact_vs_dtwa} shows the excellent agreement of the dTWA calculation with the exact solution for crystals with $M=4$ and $M=5$ rings, confirming the validity of the dTWA technique for beyond-mean-field calculations in this work.

\begin{figure}[!tb]
    \centering
    \includegraphics[width=0.4\textwidth]{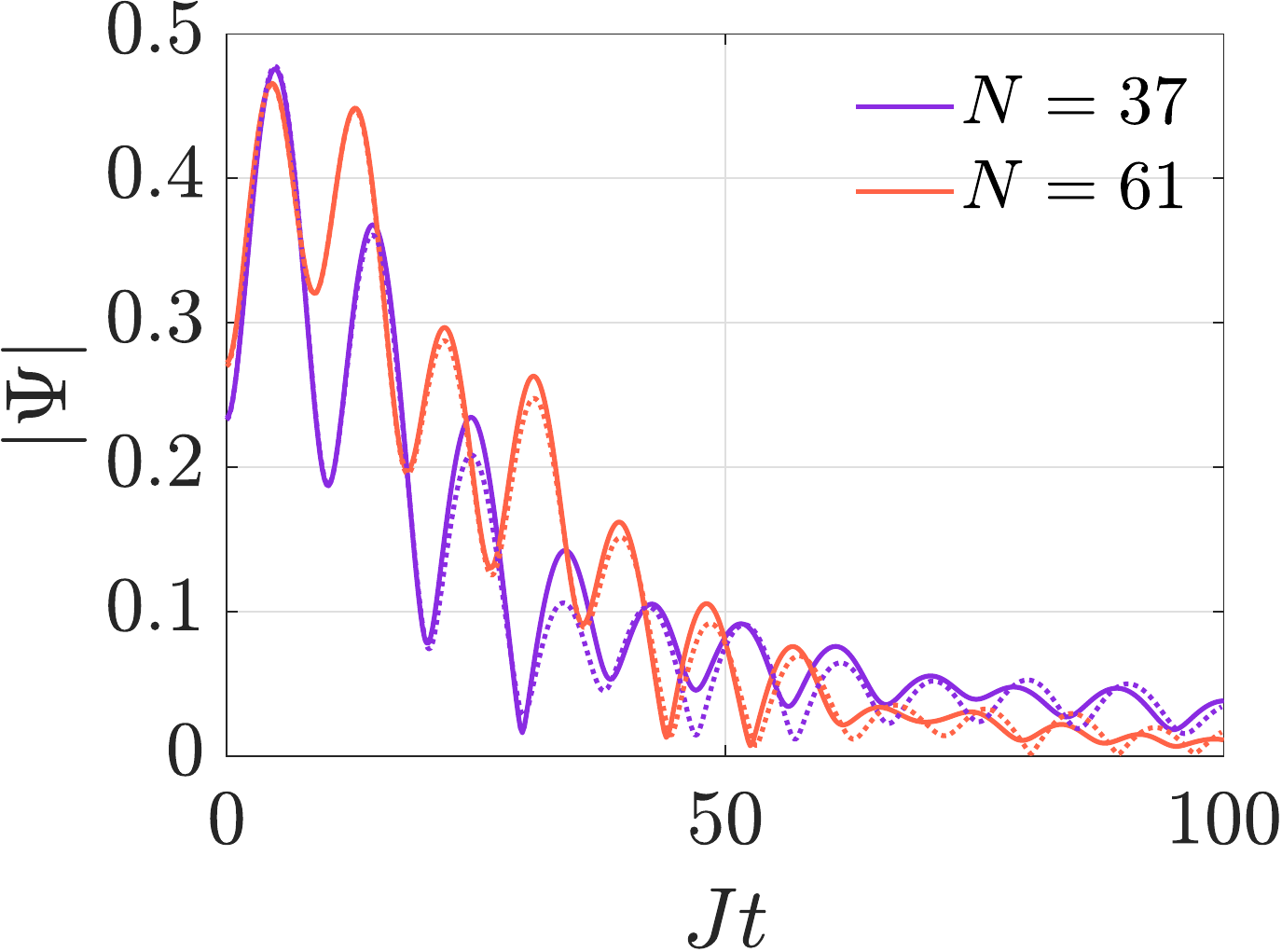}
    \caption{Benchmarking the discrete truncated Wigner approximation (dTWA) method. Solid lines are computed by numerical propagation of the Schrodinger equation while the dotted lines are obtained using the dTWA method. Here, we assume that the crystal is made of perfectly concentric rings of ions and the spin state is initialized in a BCS-like state [top panel of Fig.~\ref{fig:fig2}(a)]. Here, we have used the ratio $K/J=1$.}
    \label{appfig:exact_vs_dtwa}
\end{figure}

\section{\label{appsec:winding}Computation of winding numbers}

In continuous 2D space, the winding number of a unit vector field $\hat{\mbf{u}}(x,y)$ is defined as the surface integral 
\begin{eqnarray}
W = \frac{1}{4\pi}\int dx \; dy \; \hat{\mbf{u}} \cdot \left(\frac{d\hat{\mbf{u}}}{d x} \times \frac{d\hat{\mbf{u}}}{d y} \right).    
\end{eqnarray}
In the crystal, the winding number calculation must be carried out on a lattice with discrete sites. Here, the prescription is to identify triplets of neighbors by introducing a triangulation of the crystal lattice as shown in Fig.~\ref{appfig:delaunay}~\cite{muller2018thesis}. We use the Delaunay triangulation, wherein triangles are formed between neighboring triplets in such a way that no vertex of the crystal lies inside the circumcircle of each triangle. For each triangle, we label the vertices $A,B,C$ such that the directed edges  give rise to a face normal pointing upward from the crystal plane, i.e. $\overrightarrow{AB}\times \overrightarrow{BC} \parallel \hat{\mathbf{e}}_z$. Having identified such ordered triplets of vertices, a solid angle $\Omega_{ABC}$ is introduced for each triangle, defined as 
\begin{eqnarray}
\tan\left(\frac{\Omega_{ABC}}{2}\right) = \frac{\hat{\mbf{u}}_A\cdot \left(\hat{\mbf{u}}_B \times \hat{\mbf{u}}_C\right)}{1 + \hat{\mbf{u}}_A\cdot\hat{\mbf{u}}_B + \hat{\mbf{u}}_B\cdot\hat{\mbf{u}}_C + \hat{\mbf{u}}_C\cdot\hat{\mbf{u}}_A}.
\end{eqnarray}
The winding number on the discrete lattice is obtained by summing the solid angle $\Omega_{ABC}$ over all triangles $\Delta_{ABC}$ of the triangulation:
\begin{eqnarray}
W = \frac{1}{4\pi} \sum_{\Delta_{ABC}}\Omega_{ABC}.
\end{eqnarray}

\begin{figure}[!tb]
    \centering
    \includegraphics[width=0.5\textwidth]{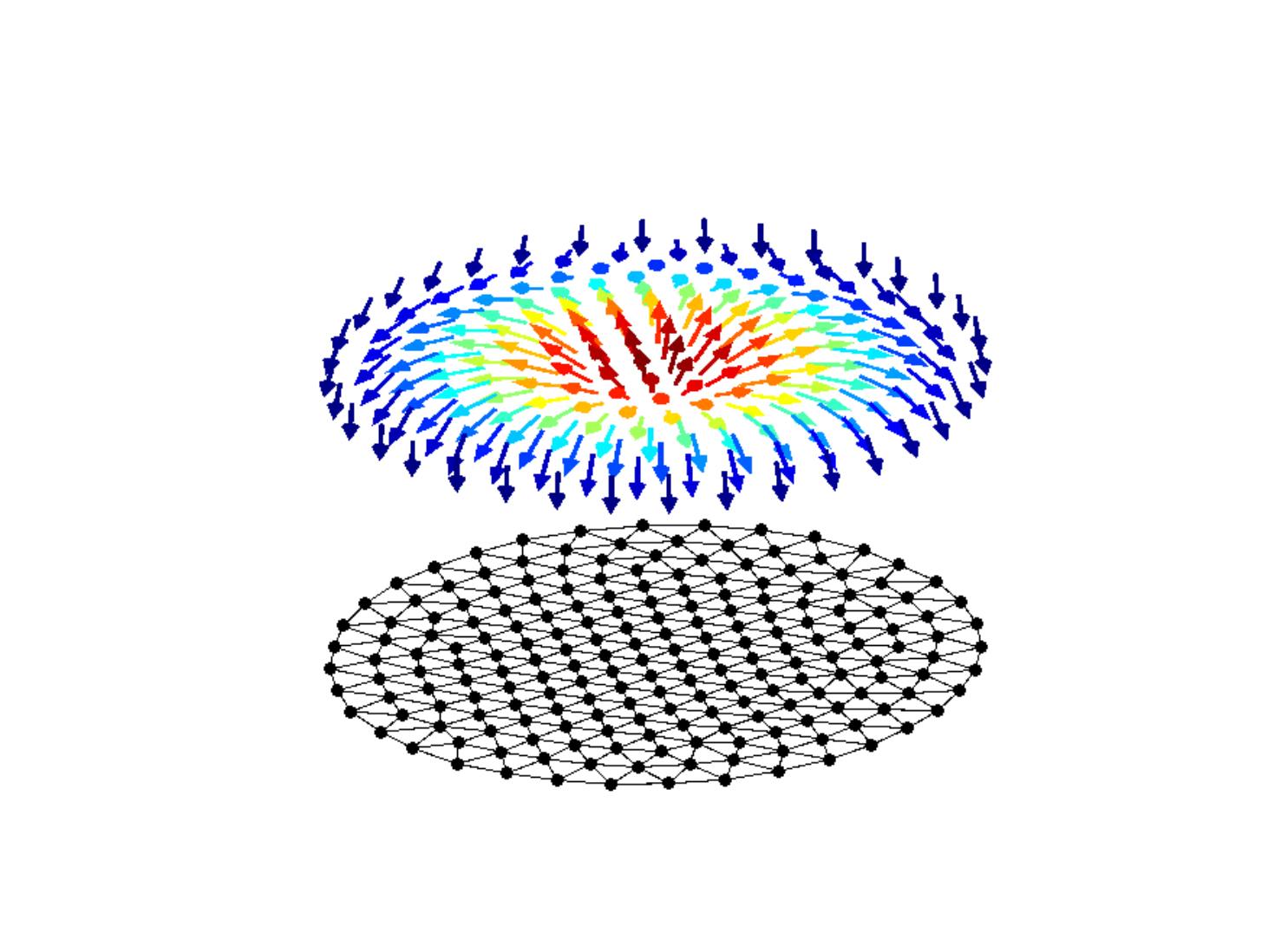}
    \caption{Demonstration of Delaunay triangulation for a crystal with 200 ions. The triangulation is used to identify triplets of neighbors on the discrete lattice for computing the winding number of the spin texture ($Q$) or that of the effective magnetic field texture ($W$).}
    \label{appfig:delaunay}
\end{figure}

In the present work, we call the winding number $Q$ if the vector field is taken as the spin texture, while we label it as $W$ if the vector field is the effective magnetic field texture in the frame rotating at $2\mu_\infty$. 

\section{\label{appsec:init}State initialization}

To generate chiral BCS-like and BEC-like initial states, we take advantage of the term of the form $A\cos C \hSS_j^-$ that is available in the small-angle expansion of the ODF interaction [Eq.~(\ref{appeqn:small_ang})]. This term describes a coupling of the spins with the planar rotation without involving the drumhead c.m. mode. We assume that the ODF lasers have a tunable beam waist $w_\mathrm{ODF}$. By tuning $\mu_r = B_0+\omega_r$ and ignoring rapidly oscillating terms, the effective interaction is given by 
\begin{eqnarray}
\hat{H}_\text{init}= \sum_j \frac{\Omega_j}{2}  \left(\hSS_j^+ e^{-i\phi_j} + \hSS_j^- e^{i\phi_j} \right),
\label{appeqn:h_init}
\end{eqnarray}
where $\Omega_j = \Omega_0 \td{r}_j e^{-r_j^2/{w_\mathrm{ODF}^2}}$ with $\Omega_0 = \delta_\mathrm{AC}(\Delta k_x R)/2$ and $\td{r}_j=r_j/R$. The Hamiltonian $\hat{H}_\mathrm{init}$ describes non-interacting spins each undergoing rotation under a local magnetic field. Using the local axes introduced in Eq.~(\ref{eqn:local_axes}), we can compactly write 
\begin{eqnarray}
\hat{H}_\mathrm{init} =  \sum_j \Omega_j \hSS_j^{{\mcY_j'}}.
\end{eqnarray}
For the initialization, we assume that the beam waist of the Raman beams is much larger than the crystal radius, i.e. $w\gg R$ so that the dispersion arising from the spatial variation of the Raman beams can be neglected. 

We now describe the initialization protocols for various cases that we discuss in the Main Text. 

\subsection{\label{appsec:bcs_phase2}BCS initialization for phases I and II}

We assume that the beam waist of the ODF lasers are much larger than the crystal radius, i.e., $w_\mathrm{ODF}\gg R$. Then, $\hat{H}_\mathrm{init}$ reduces to 
\begin{eqnarray}
\hat{H}_\mathrm{init} =  \Omega_0 \sum_j  \td{r}_j \hSS_j^{{\mcY_j'}}.
\end{eqnarray}
We initialize all spins in $\ket{\uparrow}_{{\mcZ}}$ (i.e. in the rotated spin space) by an appropriate global $\pi/2$ pulse. Setting the maximum pulse area $\Omega_0 T = \pi$, the outermost spins are then rotated all the way to $\ket{\downarrow}_{{\mcZ}}$ whereas the central spin is unaffected by virtue of the dependence of the Rabi frequency on $\td{r}_j$. Data shown in Fig.~\ref{fig:fig2}(a-b) and Fig.~\ref{fig:fig4} are obtained using this initial state.

\subsection{BEC initialization for  winding number studies}

We assume that $w_\mathrm{ODF}<\sqrt{2}R$. The radial modulation of the Rabi frequency results in a maximum Rabi frequency $\Omega_\mathrm{max}$ at radius $r_\mathrm{max}$, that are respectively given by 
\begin{eqnarray}
\Omega_\mathrm{max} = \Omega_0\frac{w_\mathrm{ODF}}{R\sqrt{2e}},\; r_\mathrm{max} = \frac{w_\mathrm{ODF}}{\sqrt{2}}.
\end{eqnarray}
Therefore, for $w_\mathrm{ODF}<\sqrt{2}R$, the maximum pulse area for a fixed rotation time $T$ is experienced by a spin somewhere in the middle of the crystal. We initialize all spins in $\ket{\downarrow}_{{\mcZ}}$. For the winding number study, we set the maximum pulse area $\Omega_\mathrm{max}T = \pi$ and use a beam waist $w_\mathrm{ODF}=0.3R$, which ensures that a large number of the outer spins are negligibly rotated. This ensures  that the winding number $W$ is reasonably quantized. Data shown in Fig.~\ref{fig:fig3} are obtained using this initial state.

\subsection{BCS initialization for phase III}

Preparing the initial state for phase III requires the presence of a sharp domain wall and an order parameter of small magnitude. To obtain the domain wall, first spins are initialized in $\ket{\downarrow}_Z$ (i.e. in the unrotated spin space). An optical pumping beam selectively excites spins in a central region of radius $r_d$ (chosen to be $r_d=R/2$) to $\ket{\uparrow}_Z$. A $\pi/2$ pulse about the $-Y$-axis then respectively converts the central and outer regions to domains of $\ket{\uparrow}_{{\mcZ}}$ and $\ket{\downarrow}_{{\mcZ}}$ spins (i.e. in the rotated spin space). To initialize a small order parameter, we assume $w_\mathrm{ODF} = R/2$ and set the maximum pulse area to be $\Omega_\mathrm{max}T = 0.1\pi$, i.e. the spin rotation is through rather small angles. Furthermore, since the spins in the different domains start in opposite orientations, a partial cancellation occurs that further decreases the magnitude of the initial order parameter. Data shown in Fig.~\ref{fig:fig2}(c) are obtained using this initial state.

\end{appendix}

\end{widetext}



%

\end{document}